\documentclass[AMA,LATO1COL]{WileyNJD-v2-tmp}
\usepackage{moreverb}

\usepackage{csquotes}
\usepackage{makecell}
\usepackage[export]{adjustbox}
\usepackage{multirow}
\usepackage{tabularx}
    \newcolumntype{L}{>{\raggedright\arraybackslash}X}
\usepackage{subcaption}

\hypersetup{hidelinks}

\usepackage{listings}

\usepackage[final]{changes}
\definechangesauthor[color=orange]{HA}

\newcommand\BibTeX{{\rmfamily B\kern-.05em \textsc{i\kern-.025em b}\kern-.08em
T\kern-.1667em\lower.7ex\hbox{E}\kern-.125emX}}

\articletype{Article Type}%

\received{02 March 2023}
\revised{31 May 2024}
\accepted{05 October 2024}
\doi{10.1002/smr.2738}

\begin{document}

\title{Multi-Language Detection of Design Pattern Instances}

\author[1,2]{Hugo Andrade*}
\author[1,2]{João Bispo}
\author[1,2]{Filipe F. Correia}

\authormark{HUGO ANDRADE \textsc{et al}}

\address[1]{\orgdiv{Faculty of Engineering}, \orgname{University of Porto}, \orgaddress{\state{Porto}, \country{Portugal}}}

\address[2]{\orgname{INESC TEC}, \orgaddress{\state{Porto}, \country{Portugal}}}

\corres{*Hugo Andrade. \email{up201003007@up.pt}}
% \corres{**João Bispo. \email{jbispo@fe.up.pt}}
% \corres{***Filipe F. Correia. \email{filipe.correia@fe.up.pt}}

% \presentaddress{Present address}

\abstract[Abstract]{
\added{Code comprehension is} often supported by source code analysis tools which provide more abstract views over software systems, such as those detecting design patterns. These tools encompass analysis of source code and ensuing extraction of relevant information.
However, the analysis of the source code is often specific to the target programming language.

We propose DP-LARA, a multi-language pattern detection tool that uses the multi-language capability of the LARA framework to support finding pattern instances in a code base. 
LARA provides a virtual AST, which is common to multiple OOP programming languages, and DP-LARA then performs code analysis of detecting pattern instances on this abstract representation.

We evaluate the detection performance and consistency of DP-LARA with a few software projects. Results show that a multi-language approach does not compromise detection performance, and DP-LARA is consistent across the languages we tested it for (\textit{i.e.}, Java and C/C++).
Moreover, by providing a virtual AST as the abstract representation, we believe to have decreased the effort of extending the tool to new programming languages and maintaining existing ones.}


\maketitle

\section{Introduction}\label{intro}

Design Patterns (DPs) provide templates of general solutions to common software development problems. In addition to being proven and tested solutions that help speed up and make code more robust, the broad use and familiarity with DPs provide a common language that ultimately helps code readability and comprehension in software maintenance processes, and helps communication within a software team~\cite{zhang2012effectiveness,lemos2024live}. Therefore, it is not surprising that code analysis tools have been used to automatically detect the presence of design patterns in code.

\footnotetext{\textbf{Abbreviations:} DP, Design Pattern; DPD, Design Pattern Detection; AST, Abstract Syntax Tree; OO, Object Oriented }

However, while software projects are becoming increasingly complex, often encompassing multiple languages and technologies, it may be limiting that these tools are specific to a single programming language. When employing them, teams and organizations would need to use various design pattern detection (DPD) algorithms, which must be maintained throughout their usage. Maintaining a large tool suite for the same task is redundant work for developers, which ultimately hampers their adoption. In addition, different tools for the same task would lead to inconsistent results and different output formats, since each may have its own interpretation of DPs and follow a different detection approach.

Multi-language approaches would help deal with these problems, as they implement a sole detection algorithm using a language-agnostic representation of the system. However, this transformation step of source code to meta-representation is often complex and is particularly laborious when extending the tool to add support to other programming languages, as they need to re-implement entire modules.
Therefore, we propose DP-LARA, a multi-language framework for DPD that merges a language-specific detection approach identified in the existing literature with the multi-language capabilities of LARA~\cite{LARA}, a framework for source-to-source compilation that can reduce the effort to extend to new programming languages.

We start with the literature review on DPD approaches in Section~\ref{related-work}, then we describe DP-LARA in Section~\ref{DP-LARA}. Finally, we detail a series of experiments ran on DP-LARA in Section~\ref{evaluation}, discuss the results in Section~\ref{discussion}, and provide conclusions and future work in Section~\ref{conclusion}.

\section{Related Work}\label{related-work}

Detection of DPs from source code has been a topic explored for a few decades and encompasses very different approaches and results. An extensive review by Hadis \& Seyed~\cite{DPDReview} studied the different techniques and the data representation most commonly used, their limitations and advantages, and discussed the open issues in this field. Among their conclusions, the authors ``expect a move towards language-independent approaches''.

%
% RULES SATISFACTION
% 
Several language-specific tools follow a rules-satisfaction approach, where DPs are described as collections of conditions that must be satisfied. DP-CORE~\cite{DPCORE} follows this approach. It uses a UML-based language for representing patterns, and extracts information from the Abstract Syntax Tree (AST), which is then mapped to this language. 
\added[id=HA]{Other rules-based approaches include PINOT~\cite{PINOT}, FUJABA~\cite{FUJABA}, and tool by Heuzeroth \textit{et al.}~\cite{Automatic-AST}, which also use AST or variants as the intermediate representation of the system. However, the definitions of DPs are ingrained in the algorithms, thus hampering their flexibility to detect other DPs without changes in the codebase. Similar approaches use meta-patterns~\cite{pree1995design} (a finite set of abstractions of inter-class relationships) as an intermediate representation for the rule-satisfaction detection algorithm. JFREEDOM~\cite{JFREEDOM} and the tool by Hayashi~\textit{et al.}\cite{MetaPatterns} are examples of this approach, but they have the same issue of ingrained definitions of DPs, and the use of meta-patterns requires language-specific algorithms for their detection.}

%
% MACHINE LEARNING
% 
Machine learning is also used in DPD tools. The tool by Thaller \textit{et al.}~\cite{FeatureMaps} detects micro-structures and normalizes them into feature maps for the learning step, but also requires language-specific algorithms for their detection. DPDf~\cite{Feature-based} extracts features directly from source code with syntactic and semantic analysis, which makes the results very dependent on the language syntax and code-writing style of the developers. Chihada \textit{et al.}~\cite{ML-Classification} and Mhawish \textit{et al.}~\cite{ML-TreeBaseAlgorithm} extract Object-Oriented (OO) metrics from each class and use different models whether interpretability of the prediction model is a concern or not, but, as all machine-learning-based approaches, they require huge amounts of labelled data and retraining of the model every time there is a change, which evidently hampers the flexibility of their approach.

%
% GRAPH-BASED
%
Graph-based approaches for DPD transform the source code and the definitions of DPs into graphs and, through graph-matching algorithms, detect the respective instances. Gupta \textit{et al.} and explored several graph matching techniques~\cite{Graph-InexactMatching, Graph-Greedy, %Graph-SubgraphIsomorphism, 
Graph-CrossCorrelation, Graph-DPDetect, Graph-Hybrid}, Tsantalis \textit{et al.}\cite{Graph-Similarity} proposed a similarity scoring algorithm, and Dong \textit{et al.}~\cite{Graph-TemplateMatching} complemented its shortcomings with a template matching algorithm. MARPLE~\cite{MARPLE%, MARPLE-ML
} combines graph-matching, when detecting candidate instances, with ML, by then applying a neural network to refine the detection. Graph-based approaches have comparable flexibility as rule-satisfaction, because DPs can be described separately in the graph-based representation language of the tool. Even though graph-based approaches require an additional step to transform static information into graphs, it is often straightforward unless the graph-based model is complex. However, these tools were not publicly available.

%
% MULTI_DP
% 
Some multi-language approaches have been proposed. nMARPLE~\cite{MARPLE-NET} provides multi-language support by re-implementing in another language the first module of their pipeline --- the detection of micro-structures. The tool by Fabry \& Mens~\cite{FABRY200421} transforms source code to a meta logic programming language, and reasons about the detection of DPs at this level by a series of queries and verification of logic facts. Nagy \& Kovari~\cite{NEUTRAL} use a criterion-based detection algorithm on an instance of an OO data model where language-specific processors are responsible for producing an instance of the data model. Similarly, CrocoPat~\cite{CrocoPat} and DPF~\cite{Graph-DSL} also define a meta-model and use different graph-based detection algorithms.

%
% DISCUSSION
%

In sum, we observe that there are several approaches to DPD. Even though most are language-specific, the detection step is not performed directly on source code but on an intermediate representation of the system. This is a critical point for adding multi-language capabilities to a detection tool: how strongly bound the intermediate representation is to the language of the program, and how is the translation of the source code to the intermediate representation done. Some multi-language approaches re-implement entire modules of previous language-specific approaches, while others define a meta-representation. The detection methodology influences the complexity of the intermediate representation that it needs, which affects the effort required for the translation step; the more complex it is, the more it hinders the tool's development, maintenance, and future improvements.

The flexibility of multi-language DPD approaches to add support to new programming languages highly depends on the effort required in the translation step. AST provides a close representation of the source code and most of the (static) information is still preserved. In addition, AST analysis is the basis of most of the approaches, and it is often the sole source of information for rule-based approaches. Therefore, injecting multi-language support at this level by providing a language-agnostic AST should be less complex and require less effort, since most of the work consists of mapping similar concepts. It should also preserve the most information. Finally, rule-based approaches have the flexibility of adding new definitions of DPs by describing them separately in the representation language. 

This work proposes a multi-language approach that has both benefits and which we name DP-LARA.

\section{DP-LARA}\label{DP-LARA}

DP-LARA is a re-implementation of an existing and proven DPD tool, DP-CORE~\cite{DPCORE}, in a multi-language framework for source-code analysis, LARA~\cite{LARA}. During this work, we extended the multi-language capabilities of LARA, so that we could extract the information required for the DPD algorithm.

Developing a novel DPD algorithm for the LARA framework was an option that was excluded after reviewing the literature. Since many capable approaches have been developed over the past few decades, this work decided \added[id=HA]{that it would be a greater contribution to this field} to select an existing one and add multi-language capabilities.

\subsection{Goals and requirements}

The primary goal was to develop a multi-language tool for DPD based on LARA. LARA is an Aspect-Oriented Programming framework with multi-language properties since it allows the development of aspects that can be applied to different programming languages. 

An additional goal was to develop a flexible approach that would encourage the continuous evolution of DP-LARA with reduced effort for developers. Most tools did not consider preceding work other than comparison of results, likely because the code-base was not (publicly) available. We consider that it hampers the evolution of this field and resulted in the creation but seldomly the evolution of DPD tools.

Following these goals, we defined the following requirements. The tool should have good detection results, support multiple languages, a flexible mechanism for adding new definitions of DP (or modifying existing ones), and reduced effort for adding new languages and maintaining existing ones. We consider these requirements were satisfied with the selection of DP-CORE and the architecture of LARA. 

\subsection{UML-based Concepts}

The elected direction was to select an existing and proven DPD tool and readjust it to be used with LARA. DP-CORE~\cite{DPCORE} satisfied these requirements, since the detection algorithm uses a UML-based representation language, is well documented and (publicly) available, and the definitions of DPs are easily customized. 
The definitions are described as collections of objects with abstraction levels and relationships between objects, as illustrated in Figure~\ref{figure:dpcore-pattern-file}. To add support to other DPs, the user simply needs to describe them in this UML-based representation language, which is further explained in the DP-CORE paper~\cite{DPCORE}.

\begin{figure}[htbp]
\centering
\fontsize{8}{10}\selectfont
\begin{tabular}{l} 
 \hline
\texttt{Observer}\\
\texttt{A Normal Concrete Observer}\\
\texttt{B Abstracted Observer}\\
\texttt{C Any Subject}\\
\texttt{End\_Members}\\
\texttt{A inherits B}\\
\texttt{A calls C}\\
\texttt{C references B}\\
\texttt{End\_Connections}\\
 \hline
\end{tabular}
\caption{File used by DP-CORE to define the \textit{Abstract Factory} design pattern, showing rules expressed in a UML-based representation language.}
\label{figure:dpcore-pattern-file}
\end{figure}

DP-CORE uses the AST as an intermediate representation of the source code and maps the extracted information into the representation language. It uses core concepts of OO Programming, therefore they can be applied to multiple OO languages that share the same concepts. DP-CORE defined five Abstraction Types originally for Java, but they were slightly altered to accommodate other OO languages.
\texttt{Normal} refers to a simple non-abstract class that can be instantiated, \texttt{Interface} corresponds to a class with only declarations and no state, and \texttt{Abstract} corresponds to classes that cannot be instantiated but can have fields and abstract methods. Finally, \texttt{Abstracted} is either one of the types \texttt{Abstract} and \texttt{Interface}, and \texttt{Any} is any of the above-mentioned types. In addition, the connections between objects are extracted as directional relationships. Table~\ref{table:dev:dp-core-connections} summarizes the six connection types defined by DP-CORE.

\begin{table}[htbp]
\begin{center}
\caption{DP-CORE Connection types and their descriptions.}
% \small
\fontsize{8}{10}\selectfont
\setlength{\tabcolsep}{6pt} % Default value: 6pt
\renewcommand{\arraystretch}{1.4} % Default value: 1
\begin{tabular}{l l} 
 \hline 
 Connection Type & Description \\ [0.5ex]
 \hline
 \texttt{A inherits B}   
 & Class A inherits or implements class B or class A realizes interface B \\ 
 \texttt{A has B}        
 & Class A has one or more objects of type B  \\ 
 \texttt{A references B} 
 & A method of class A has as parameter an object of type B \\ 
 \texttt{A creates B}    
 & Class A creates an object of type class B \\ 
 \texttt{A uses B}       
 & A method of class A returns an object of type B  \\ 
 \texttt{A calls B}      
 & A method of class A calls a method of class B  \\ 
 \hline
\end{tabular}
\label{table:dev:dp-core-connections}
\end{center}
\end{table}

Finally, the actual detection is performed by matching the description of the DP with the extracted information (abstract types and connections) of the objects of the source code. This matching step consists of a recursive algorithm, and since both inputs are defined in the UML-based representation language, this step is language-independent.

\subsection{Architecture and design}\label{DP-LARA:architecture}

%
% LARA ARCHITECTURE
% 
LARA's architecture allows multi-language capabilities. For each specific language (or set of languages) it supports, there is a corresponding LARA compiler and language specification. They are responsible for bridging the gap between the agnostic LARA aspects and the target language. Thus, LARA provides a common interface between languages, based on a virtual AST. Currently, there are LARA compilers for the C/C++ (Clava~\cite{bispo2020clava}), Java (Kadabra~\cite{carvalho2023dsl}), JavaScript (Jackdaw), and MATLAB (MATISSE~\cite{reis2020compilation}) programming languages, whose codebase can be found in the SPeCS repository\footnote{\href{https://github.com/specs-feup}{https://github.com/specs-feup}}, with a new LARA compiler for Smali (i.e., Android assembly) in development.

\begin{figure}[htbp]
\centerline{\includegraphics[width=0.75\linewidth]{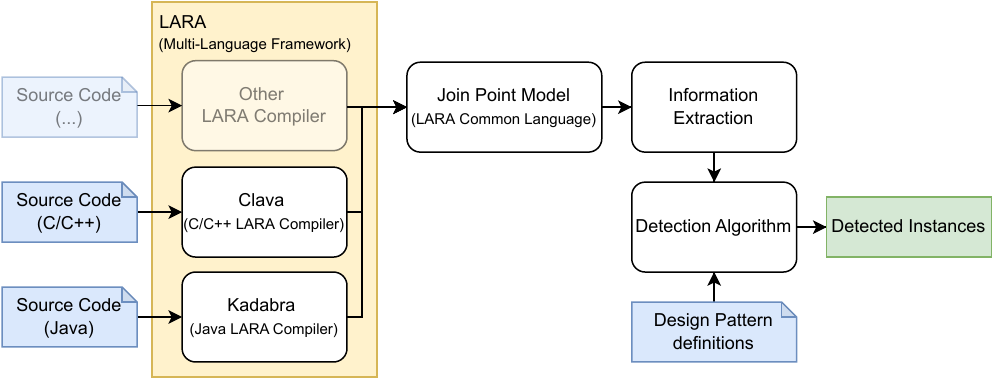}}
% \centerline{\includegraphics[width=0.75\linewidth]{images/Pipeline-LARA.jpg}}
\caption{DP-LARA Architecture.}
\label{figure:dp-lara-architecture}
\end{figure}

However, the different nomenclatures and language specifications hinder the full agnosticism of LARA aspects. For instance, a \texttt{class}, which is a common concept of OOP languages, has different nomenclatures in each LARA compiler --- \texttt{class} in Clava (C/C++) and Kadrabra (Java), and \texttt{classDeclaration} Jackdraw (JavaScript). Another common concept in OOP is inheritance, where classes can extend other classes --- Clava uses \texttt{bases} nomenclature and returns an array of \texttt{class}, but since Java and JavaScript use Single Inheritance, their nomenclature attribute is \texttt{superClass} and returns a single value. Therefore, to fully obtain agnosticism, a common language specification with OO concepts was defined --- the \textbf{LARA Common Language}~\cite{LARAM} --- and compliant LARA compilers must implement it. The overall architecture of the tool is in Figure~\ref{figure:dp-lara-architecture}.

%
%  + List + Architecture Image
% 

\begin{figure*}[htbp]
\centerline{\includegraphics[width=1\linewidth]{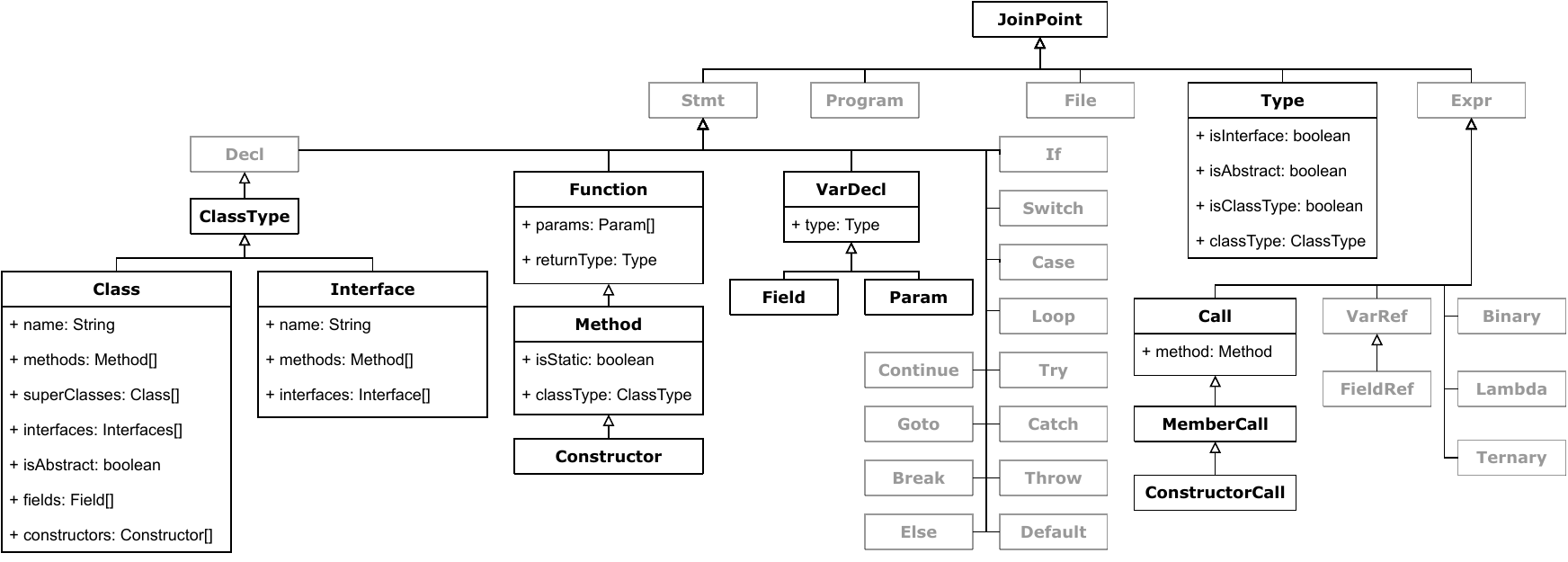}}
% \centerline{\includegraphics[width=1\linewidth]{images/LARAM.jpg}}
\caption{LARA Common Language Join Point Model. 
The join points critical in the context of DP-LARA are non-faded and only required attributes are listed.}
\label{figure:lara-common-model}
\end{figure*}

Its implementation follows two steps: (1) definition of the Common Language Specification, which is a common set of join points and attributes that will be shared between programming languages; (2) mapping the join points to the nodes of the AST of each LARA Compiler, and provide the implementation for their attributes. Figure~\ref{figure:lara-common-model} illustrates the LARA Common Language join point model, highlighting the join points and attributes required for the extraction of abstraction types and connections. 

Finally, the communication between the LARA Common Language and the AST of the LARA Compilers is handled by the \texttt{AstMethods}, a Java Interface which acts as a bridge between compilers and the LARA aspects. LARA aspects communicate with the AST of the compilers using this centralized class. On the other hand, LARA Compilers must implement this interface. It has three types of methods:

\begin{itemize}

    \item \textbf{AST transverse methods}: methods to navigate the ASTs: \texttt{getRoot}, \texttt{getParent}, \texttt{getDescendants}, \texttt{getChildren},  \texttt{getScopeChildren}.

    \item \textbf{Mapping method}: method responsible to map the AST nodes to join points of the Common Language Specification: \texttt{getJoinPointName}.

    \item \textbf{Compiler Utility methods}: method that converts an AST node to the join point of the native language specification of the LARA Compiler: \texttt{toJavaJoinPoint}. As stated before, some issues can be simple differences in nomenclature or return formats, and this method helps reuse the already implemented language-specific functionality.

\end{itemize}

%
% DECISION OF PORTING
%
% MODIFICATIONS TO CONCEPTS FOR MULTI-LANGUAGE
%
The UML-based concepts of DP-CORE were migrated to LARA, and the steps for extraction of information and detection matching algorithm were re-implemented in LARA. 

DP-LARA supports C/C++ and Java, which means that the respective LARA Compilers are compliant with the updated LARA Common Language. In doing so, these languages' peculiarities had to be handled in a multi-language approach. For Java, since it uses Single Inheritance, the output formats of some attributes were normalized to arrays. For C/C++, classes are declared and defined in different files, and the occurrence of duplicates was avoided by giving preference to implementations and discarding other occurrences. Moreover, C/C++ does not have a directly correspondent reference type to \texttt{Interface}, so the altered definitions of abstraction types needed to be verified when mapping language-specific AST nodes to the common language join points. 

Finally, the information for the detection algorithm was extracted from the join point model, and the steps required for each are detailed below:

\begin{itemize} 
\setlength\itemsep{-0em}

\item\textbf{\texttt{Abstract Types}}: 
All class objects are represented by the join point \texttt{ClassType}. Those with abstract type \texttt{Interface} extend the corresponding join point. The types \texttt{Abstract} and \texttt{Normal} are distinguished by the boolean attribute \texttt{isAbstract} of the \texttt{Class} join point.

\item\textbf{\texttt{Inherits} connection}: For the \texttt{Interface} join point, the attribute \texttt{interfaces} returns an array of the interfaces that the class represented by that join point implements. The \texttt{Class} join point has an identical attribute, in addition to \texttt{superClasses}.

\item\textbf{\texttt{Has} connection}: \texttt{Class} join point has the \texttt{fields} attribute, which returns an array of \texttt{Field} join point. It inherits the attribute \texttt{type} of \texttt{VarDecl} join point, which is used to extract the type of the class of each field.

\item\textbf{\texttt{References} connection}: \texttt{ClassType} join point has the \texttt{methods} attribute, which returns an array of \texttt{Method} join points. They inherit from \texttt{Function} which has an attribute that returns the parameters as a \texttt{Param} join point. \texttt{Param} inherits the attribute \texttt{type}, which is used to extract the type of the class.

\item\textbf{\texttt{Uses} connection}: Similarly to the previous connection type, the \texttt{Method} join point inherits the \texttt{returnType} attribute from \texttt{Function}, which is a \texttt{Type} and is used to extract the type of the class.

\item\textbf{\texttt{Creates} connection}: the ``calls'' to constructors are queried within the class objects, and they are represented by the \texttt{ConstructorCall} join point. It inherits from the \texttt{Call} join point which has an attribute of type \texttt{Method}, and in turn \texttt{Method} has a \texttt{classType} attribute to inform the type of the class to where that method belongs.

\item\textbf{\texttt{Uses} connection}: Similarly to the previous connection type, the ``calls'' are queried within the class objects, and they are represented by the \texttt{Call} join point. Then, one arrives at the type of the class in a similar process.

\end{itemize}

\subsection{Usage}

DP-LARA\footnote{\textbf{DP-LARA:} \url{https://github.com/specs-feup/lara-framework/tree/master/LaraCommonLanguageApi/src-lara/lcl/patterns}} is part of LARAM, a library written in LARA that uses the LARA Common Language, whose original objective was multi-language metrics extraction~\cite{LARAM}. The detection algorithm requires two inputs: the definitions of DPs in the language representation defined in the previous section and the code base that will be analyzed. DP-LARA can be used within the LARA framework by simply importing the LARAM library. Several weavers have been developed, each for a specific language (or set of languages) --- \textit{e.g.}, Clava for C/C++, Kadabra for Java --- and the language of the code-base to be analyzed must match the appropriate weaver. 

Potential applications of DP-LARA can include its integration in plugins for IDEs or the creation of a single console application that includes several LARA compilers that support LARAM, which can detect the set of languages that a code base has and analyze the code with the appropriate compilers.

\added[id=HA]{DP-LARA can analyze non-compilable source code, which is an option the developer might choose. Projects usually have multiple dependencies of large libraries, which have a multitude of classes that are of low interest and low usefulness to the project. Parsing projects with all their dependencies would generate huge amounts of information, which would exponentially degrade the performance of the algorithm. Therefore, we chose a different approach that parses the non-compiled source code of projects even if some dependencies are missing and the project itself is still not compilable. We consider it a reasonable compromise because the loss of some information of (frequently) low-interest library code is offset by the increase in performance due to the considerable decrease in information space.}

\section{Evaluation} \label{evaluation}

We performed a series of validation studies that aim to validate that a multi-language approach for code analysis (\textit{Ic}. the detection of design pattern instances) has good performance, is consistent in the multiple languages it supports, and is flexible by requiring less effort to extend support to new languages. Throughout this series of experiments, we use the definitions of DPs defined in the DP-CORE project\footnote{\textbf{DP-CORE@Github:} \url{https://github.com/AuthEceSoftEng/DP-CORE}} at this commit\footnote{\textbf{DP-CORE@dba4dcb's Patterns:}\url{https://github.com/AuthEceSoftEng/DP-CORE/tree/dba4dcb/patterns}}, which are \textsc{Abstract Factory}, \textsc{Bridge}, \textsc{Builder}, \textsc{Command}, \textsc{Observer} and \textsc{Visitor}. All the repositories and LARA aspects mentioned and used throughout this work are available in a code repository\footnote{\textbf{DP-LARA's config files:} \url{https://github.com/specs-feup/lara-dpdetection}} for replication, as well as instructions in the README.md file to build, run and test the experiments. 

Therefore, we have formulated the following Research Questions:

\begin{itemize}

    \item[] \textbf{RQ 1}: \textit{Is the multi-language approach as precise as language-specific DPD?} --- Having multi-language capability should not hinder the precision and recall metrics of the design pattern approach. It should have equal or better results than existing language-specific tools. \added[id=HA]{This research question is addressed in Sections~\ref{evaluation:detection-range}~and~\ref{evaluation:detection-performance}.}

    We do a comparative analysis with DP-CORE, the selected language-specific DPD algorithm. Even though DP-LARA does not use the original detection algorithm, efforts were made to migrate it unalterably to LARA. Thus, the same information fed into either one of the implementations should result in approximately the same output. The difference between the two implementations is in the extraction of the information from the source code, which is where the multi-language property replaced the language-bound module. Thus, two different experiments are performed. First, we assess the \textbf{detection range} of DP-LARA in all the languages it supports by checking which types of design patterns it is able to detect. We used Java code snippets of design patterns extracted from the DP-CORE project at this commit\footnote{\textbf{DP-CORE@dba4dcb's Code Snippets (Java):} \url{https://github.com/AuthEceSoftEng/DP-CORE/tree/dba4dcb/examples}}, and equivalent C++ code snippets from \emph{Refactoring.Guru}\footnote{\textbf{Refactoring.Guru:} \href{https://refactoring.guru}{https://refactoring.guru}}, an informative website about many programming topics including GoF design patterns, publicly available in Github\footnote{\textbf{Refactoring.Guru-C++@846eb95's Code Snipets (C++):} \url{https://github.com/RefactoringGuru/design-patterns-cpp/tree/846eb95}}. Then, we compare DP-LARA's \textbf{detection performance} in large projects against DP-CORE.
    We used ten Java open-source projects: three from the DP-CORE original publication~\cite{DPCORE} (JHotDraw 6.0b1, Java AWT 1.3, and Apache Ant 1.6.2), allowing direct comparison with the published results; the remaining seven from P-MARt~\cite{P-MARt}, the repository of pattern-like micro-structures manually collected and classified by experts (QuickUML 2001, Lexi 0.1.1-alpha, JRefactory 2.6.24, JUnit 3.7, MapperXML 1.9.7, Nutch 0.4 and PMD), whose results can be used to compare DP-LARA against other detection tools in previous and future works.
    Experiments with larger code bases were only carried out with Java projects because we were able to obtain DP-CORE, and thus compare. For C++, detection tools (source code or executables) of C++ systems were not publicly available nor obtainable, and the projects they used in their evaluation were not properly labeled, circumstances that hinder a meaningful comparison. In summary, since we are not proposing a novel detection algorithm, we found it relevant to show that our implementation of DP-LARA provides at least equivalent results to those of DP-CORE.

    \item[] \textbf{RQ 2}: \textit{Does the multi-language DPD approach provide consistent results across the target languages it supports?} --- To fully support multi-language, the tool should have equal outcomes and results in projects equally implemented in several languages, despite the specificities that each programming language has. \added[id=HA]{This research question is addressed in Section~\ref{evaluation:detection-consistency}.}
    
    The goal in this experiment is to analyze a project implemented in Java and C++, expecting DP-LARA to have similar results in both languages. The projects analyzed by DP-LARA were CppUnit and JUnit. CppUnit was ported from JUnit by an independent developer, and the analysis of their source control revealed that they were the most similar close to the time of porting, before they followed independent and divergent paths. In addition, JUnit provides information regarding the DPs that it uses\footnote{\url{https://web.archive.org/web/20221128004056/https://junit.sourceforge.net/doc/cookstour/cookstour.htm}}. The commits close to the time of porting were used, and the version used for Junit (v3.4) is available here\footnote{\url{https://sourceforge.net/projects/junit/files/junit/3.4/}} while for CppUnit here\footnote{\url{https://sourceforge.net/p/cppunit/code/4/tree/}}. In addition, a more recent version of CppUnit was also analyzed and is available here CppUnit (1.12)~\footnote{\url{https://sourceforge.net/p/cppunit/code/606/}}.

    \item[] \textbf{RQ 3}: \textit{Is the effort low for extending the multi-language approach to support new OO languages?} --- Extending the tool to support new object-oriented programming languages should be straightforward and with limited effort.  \added[id=HA]{This research question is addressed in Sections~\ref{evaluation:extensibility}.}

    DP-LARA is compared with other multi-language design pattern detection approaches found in the literature. Measuring the effort is not as straightforward as evaluating other properties of the system. Thus, we combined several aspects that are relevant when developers are extending a multi-language framework: the number of critical locations where the developer needs to work on, the estimated number of lines of code that need to be implemented, and the presence of code generators and/or helper methods. However, since the projects of the other multi-language detection approaches were not obtained, the effort is inferred from the published documentation.
   
\end{itemize}

\subsection{Detection Range}\label{evaluation:detection-range}

The detection range of DP-LARA is evaluated by assessing the types of design patterns that it is able to detect. DP-CORE is analyzed under the same conditions to serve as a baseline and to verify that adding multi-language capabilities did not compromise the detection range of the algorithm. The results with Java code snippets are summarized in Table~\ref{table:eval:detection-capability-java}.

\begin{table*}[htbp]
\begin{center}
\fontsize{8}{10}\selectfont
\caption[Detection capability results, tested in Java code snippets.]{Detection capability results, tested in the Java code snippets of DP-CORE. Each design pattern was used to detect instances in all code snippets. DP-C means DP-CORE detected an instance of the Design Pattern Definition in the code snippet analyzed. DP-L means DP-LARA detected}
%\small
\setlength{\tabcolsep}{6pt} % Default value: 6pt
\renewcommand{\arraystretch}{1.4} % Default value: 1
\begin{tabular}{l | l | c  c  c  c  c  c} 
 \hline 
 \multicolumn{1}{c}{} & & \multicolumn{6}{c}{Design Pattern Definition} \\
 \cline{3-8}
  \multicolumn{1}{c}{} & & 
  \multicolumn{1}{c}{\textsc{\makecell[c]{Abstract\\Factory}}} & 
  \multicolumn{1}{c}{\textsc{Bridge}} & 
  \multicolumn{1}{c}{\textsc{Builder}} & 
  \multicolumn{1}{c}{\textsc{Command}} & 
  \multicolumn{1}{c}{\textsc{Observer}} & 
  \multicolumn{1}{c}{\textsc{Visitor}} \\ [0.5ex]
 \hline
 \multirow{8}{*}{\rotatebox{90}{Code Snippets}}
 & \textsc{\makecell[l]{Abstract\\Factory}}  & \makecell{DP-C\\DP-L} & - & - & - & - & -  \\ 
 & \textsc{Builder}            & - & \makecell{DP-C\\DP-L} & - & \makecell{DP-C\\DP-L} & - & - \\
 & \textsc{Bridge}             & - & - & \makecell{DP-C\\DP-L} & - & - & - \\
 & \textsc{Command}            & - & - & - & \makecell{DP-C\\DP-L} & - & - \\
 & \textsc{Observer}           & - & - & - & - & \makecell{DP-C\\DP-L} & - \\
 & \textsc{Visitor}            & - & - & - & - & \makecell{DP-C\\DP-L} & \makecell{DP-C\\DP-L} \\
 \hline
\end{tabular}
\label{table:eval:detection-capability-java}
\end{center}
\end{table*}

The results revealed not only that DP-LARA is able to detect all types of design patterns defined by the DP-CORE project but also that it has the exact same results as the DP-CORE detection algorithm. Both approaches detect the false positives for the \textsc{Command} and \textsc{Observer} definitions --- \textsc{Command} instances are detected in \textsc{Builder} code snippets, and \textsc{Observer} instances are detected in \textsc{Visitor} code snippets. Even though detection improvement is out of the scope of this work, one could propose different ways to improve the detection performance of the algorithm: improvement of the definitions that produced false positives, combination of detection results to exclude false positives (\textit{e.g.}, if an instance is both a \textsc{Visitor} and \textsc{Observer}, and since \textsc{Observer} instances are not detected as \textsc{Visitor}, then it can be classified as a \textsc{Visitor}), extension of the DP-CORE's UML-based representation language, or even add dynamic analysis.

Even though DP-LARA has an equal detection range as DP-CORE, this property is still unproven for languages other than Java. Since the DP-CORE project does not have code snippets in C++, and in order to avoid introducing bias when porting them from Java to C++, equivalent code snippets were extracted from \emph{Refactoring.Guru}. DP-LARA evaluated them and the results are summarized in Table~\ref{table:eval:detection-capability-cpp}.

\begin{table*}[htbp]
\begin{center}
\fontsize{8}{10}\selectfont
\caption[Detection capability results, tested in C++ code snippets.]{Detection capability results, tested with C++ code snippets from Refactoring.guru. Each design pattern was used to detect instances in all code snippets. D means DP-LARA detected an instance of the Design Pattern Definition in the code snippet analyzed.}
%\small
\setlength{\tabcolsep}{6pt} % Default value: 6pt
\renewcommand{\arraystretch}{1.4} % Default value: 1

\begin{tabular}{l | l | c  c  c  c  c  c} 
 \hline
 \multicolumn{1}{c}{} & & \multicolumn{6}{c}{Design Pattern Definition} \\
 \cline{3-8}
  \multicolumn{1}{c}{} & & 
  \multicolumn{1}{c}{\textsc{\makecell[c]{Abstract\\Factory}}} & 
  \multicolumn{1}{c}{\textsc{Bridge}} & 
  \multicolumn{1}{c}{\textsc{Builder}} & 
  \multicolumn{1}{c}{\textsc{Command}} & 
  \multicolumn{1}{c}{\textsc{Observer}} & 
  \multicolumn{1}{c}{\textsc{Visitor}} \\ [0.5ex]
 \hline
 \multirow{6}{*}{\rotatebox{90}{Code Snippets}}
 & \textsc{\makecell[l]{Abstract\\Factory}}  & D & - & - & - & - & -  \\ 
 & \textsc{Builder}            & - & D & - & D & - & - \\
 & \textsc{Bridge}             & - & - & D & - & - & - \\
 & \textsc{Command}            & - & - & - & D & - & - \\
 & \textsc{Observer}           & - & - & - & - & D & - \\
 & \textsc{Visitor}            & - & - & - & - & D & D \\
 \hline
\end{tabular}
\label{table:eval:detection-capability-cpp}
\end{center}
\end{table*}

The results demonstrate that DP-LARA has an equal detection range across the languages it supports, in this case, Java and C++. In C++, it is able to detect design pattern instances in the respective code snippets, and it also has the same false positive issues for the \textsc{Command} and \textsc{Observer} patterns.

\subsection{Detection Performance}\label{evaluation:detection-performance}

The previous experiment showed that DP-LARA has an equal detection range as DP-CORE, but it was executed under very controlled conditions with small and design-pattern-oriented code snippets. Therefore, this work proceeds with the evaluation of DP-LARA against DP-CORE with a dataset of ten Java open-source projects. Three of the projects were extracted from the DP-CORE original publication~\cite{DPCORE}. This enables direct comparison with the results DP-CORE reported. The remaining Java projects were extracted from P-MARt~\cite{P-MARt}, the repository of pattern-like micro-structures manually collected and classified by experts.

The detection performance of different approaches is often assessed with performance metrics of \textit{precision} and \textit{recall}. In order to make an unbiased comparison, the approaches need to be executed under the same conditions, \textit{i.e.}, examine the same projects with instances of the DPs detected and annotated by experts. However, the selected open-source projects did not meet these conditions, as some were not classified at all, and the ones extracted from P-MARt did not include some of the DP definitions used in this experiment. However, since DP-LARA follows the exact same principles as DP-CORE, the comparison of their detection performance is assessed by the raw detection results, whether they are correct or incorrect instances of the DPs. It is expected that DP-LARA, which adds multi-language capabilities to the original language-specific detection approach of DP-CORE, does not compromise the raw detection performance.

When analyzing the three open-source projects from DP-CORE's publication~\cite{DPCORE}, the comparative results with DP-CORE are presented in Table~\ref{table:eval:detection-results-dp-core}. 

\begin{table*}[htbp]
\begin{center}
\caption[Number of instances detected by DP-LARA and DP-CORE in projects from the publication.]{Number of Design Pattern instances detected by DP-LARA (L) and DP-CORE (C) for the JHotDraw, AWT and ANT}
%\small
\setlength{\tabcolsep}{6pt} % Default value: 6pt
\renewcommand{\arraystretch}{1.4} % Default value: 1
\begin{tabular}{l  c  c  c  c  c  c} 
 \hline
 & 
 \multicolumn{2}{c}{JHotDraw} &
 \multicolumn{2}{c}{AWT} &
 \multicolumn{2}{c}{ANT} \\
 \cline{2-7}
  & 
  \multicolumn{1}{c}{C} & 
  \multicolumn{1}{c}{L} & 
  \multicolumn{1}{c}{C} & 
  \multicolumn{1}{c}{L} & 
  \multicolumn{1}{c}{C} & 
  \multicolumn{1}{c}{L} \\ [0.5ex]
 \hline
 \textsc{Abstract Factory} & 20 & 20 & 12 & 41 & 4  & 6  \\ 
 \textsc{Builder} & 38 & 38 & 13 & 11 & 10 & 14 \\
 \textsc{Bridge} & 46 & 46 & 37 & 37 & 15 & 15 \\
 \textsc{Command} & 2  & 2  & 12 & 14 & 81 & 98 \\
 \textsc{Observer} & 10 & 11 & 16 & 16 & 7  & 7  \\
 \textsc{Visitor} & 0  & 0  & 2  & 2  & 1  & 1  \\
 \hline
\end{tabular}
\label{table:eval:detection-results-dp-core}
\end{center}
\end{table*}

The results of DP-CORE differ slightly from the original publication because their evaluation methodology was not detailed. The effort of obtaining the same version of the projects does not guarantee that, in fact, they are the same: they might be obtained from different sources, or the code base might not be complete (e.g., should the ``test'' directory be used, or just the ``src'' directory?). In addition, the publicly available version of DP-CORE might not be exactly the same that was used during the evaluation of the authors' publications---it can have improving modifications, or be a previous version than the one used. DP-CORE was also executed in this experiment to suppress this cause of variability between what was reported and the version that was used in this work.

Having DP-LARA and DP-CORE detecting design pattern instances for the same projects yielded in slightly different results. The only difference is the implementation of the detection approaches. Specifically, since the recursive detection algorithm was fully ported from Java to LARA, as supported by the previous experiments, we can assume that the differences are likely due to the step involving parsing and extracting object and connection types from the source code.

Furthermore, both DP-CORE and DP-LARA implement a merging step that assumes that when multiple classes participate in the same design pattern, only one pattern instance is detected for the same set of classes (i.e., candidate instances differing by only one member are considered part of the same design decision). However, this merging step might make some of the differences between the results of the two tools harder to interpret. For example, both tools might detect the same instance of an \textsc{Observer} design pattern, yielding similar outputs. Yet, one tool might miss one of the concrete \textsc{Observer} classes that the other detects. Such disparate results required a detailed inspection to identify their causes and explore potential solutions for the multi-language detection approach.

Finally, the P-MARt projects were also used for the evaluation of DP-LARA against DP-CORE, and the results are presented in Table~\ref{table:eval:detection-results-pmart}. Similar differences are also evident with these projects, which one might infer the same regarding their cause: differences in the parsing and extraction of objects and connections.

\newcommand\mC[1]{\makecell{#1}} % handy shortcut macro
\begin{table}[htbp]
\begin{center}
\caption[Number of instances detected by DP-CORE and DP-LARA in P-MARt projects]{Number of Design Pattern instances detected by DP-CORE (C) and DP-LARA (L) for the P-MARt projects}
%\small
\setlength{\tabcolsep}{6pt} % Default value: 6pt
\renewcommand{\arraystretch}{1.4} % Default value: 1
\begin{tabular}%x}{0.99\textwidth}%{l | c  c | c  c | c  c | c  c | c  c | c  c | c  c} 
{ l *{7}{|@{}p{0.8cm}@{} @{}p{0.8cm}@{} } }

 \hline 
 & 
 \multicolumn{2}{c|}{QuickUML} &
 \multicolumn{2}{c|}{Lexi} &
 \multicolumn{2}{c|}{JRefactory} &
 \multicolumn{2}{c|}{JUnit} &
 \multicolumn{2}{c|}{MapperXML} &
 \multicolumn{2}{c|}{Nutch} &
 \multicolumn{2}{c}{PMD}\\
 \cline{2-15}
  & 
  \multicolumn{1}{c}{C} & 
  \multicolumn{1}{c|}{L} & 
  \multicolumn{1}{c}{C} & 
  \multicolumn{1}{c|}{L} & 
  \multicolumn{1}{c}{C} & 
  \multicolumn{1}{c|}{L} & 
  \multicolumn{1}{c}{C} & 
  \multicolumn{1}{c|}{L} & 
  \multicolumn{1}{c}{C} & 
  \multicolumn{1}{c|}{L} & 
  \multicolumn{1}{c}{C} & 
  \multicolumn{1}{c|}{L} & 
  \multicolumn{1}{c}{C} & 
  \multicolumn{1}{c}{L} \\ [0.5ex]
 \hline
 \textsc{Abstract Factory} 
 &\mC{0} & \mC{1} & \mC{0} & \mC{0} & \mC{3}  & \mC{3}  & \mC{3} & \mC{2} & \mC{6} & \mC{6} & \mC{4}  & \mC{1}  & \mC{0} & \mC{2} \\ 
 \textsc{Builder} 
 & \mC{0} & \mC{0} & \mC{2} & \mC{2} & \mC{6}  & \mC{6}  & \mC{2} & \mC{2} & \mC{5} & \mC{8} & \mC{25} & \mC{56} & \mC{3} & \mC{7} \\
 \textsc{Bridge}
 & \mC{0} & \mC{0} & \mC{0} & \mC{0} & \mC{0}  & \mC{0}  & \mC{0} & \mC{0} & \mC{0} & \mC{0} & \mC{0}  & \mC{0}  & \mC{0} & \mC{1} \\
 \textsc{Command}
 & \mC{0} & \mC{0} & \mC{2} & \mC{2} & \mC{25} & \mC{24} & \mC{3} & \mC{2} & \mC{0} & \mC{1} & \mC{5}  & \mC{22} & \mC{0} & \mC{0} \\
 \textsc{Observer}
 & \mC{2} & \mC{2} & \mC{0} & \mC{0} & \mC{6}  & \mC{8}  & \mC{4} & \mC{5} & \mC{1} & \mC{1} & \mC{2}  & \mC{5}  & \mC{2} & \mC{3} \\
 \textsc{Visitor}
 & \mC{0} & \mC{0} & \mC{0} & \mC{0} & \mC{3}  & \mC{3}  & \mC{0} & \mC{0} & \mC{0} & \mC{0} & \mC{0}  & \mC{0}  & \mC{1} & \mC{1} \\
 \hline 
\end{tabular}
\label{table:eval:detection-results-pmart}
\end{center}
\end{table}

Even though the concepts for both the detection algorithm and the extraction information from the source code are kept, differences in the implementation led to the different results. There are three sources of variability: different parsing technologies were chosen, different interpretations of the key concepts (\texttt{Abstraction Type} and \texttt{Connection}), and inaccuracies in the implementations that did not account for all the languages' peculiarities. The inspection of the differences confirmed the presence of all three causes:

\begin{itemize} 
\setlength\itemsep{0em}
    
    \item \textbf{Inner Classes not properly parsed}: In DP-CORE, the parsing did not account for inner classes. During the information extraction phase, only the outer classes are transverse, and consequently, the connections are not correctly constructed because it assumes that the code inside of the inner class belongs to the outer class. On the other hand, DP-LARA not only transverses the inner classes but also attributes the proper information to them, which we assume was the intended way.

    \item \textbf{Static Method not filtered}: Design Patterns compose relationships between instances of classes. \texttt{Singleton} is the sole pattern whose code implementation comprises static methods but for very particular reasons. Therefore, this work decided to exclude static methods and fields from the information fed to the detection algorithm. However, DP-CORE does not have this rigor. 

    \item \textbf{\texttt{Calls} only on variables}: The finding of the \texttt{calls} connections in DP-CORE iterates over all the method invocations found in a class and checks if they are performed on a variable defined in that class. All the variables used in that class were previously added to an \texttt{ArrayList}, and it tries to find if there is any match with the method's \emph{invokee}. For instance, for the method declaration \texttt{foo.a()}, it looks for a variable that is named \texttt{foo}. This causes an issue when the \emph{invokee} is not a variable but the return type of another method (\textit{i.e.}, a chain of methods). DP-LARA is able to detect these connections since it considers all the methods and the original classes that they belong to.

    \item \textbf{Differences in method invocations' reference}: Another difference when extracting the \texttt{call} connections, is the reference of the method invocations. As stated in the previous topic, DP-CORE iterates over all the method invocations found in a class and checks if they are performed on a variable defined in that class. Therefore, the connection is with the class of the object referenced. In DP-LARA the ``called'' class is defined as the one where the method is implemented. This causes different detection results in inheritance situations when methods are defined in the superclass but the subclass is called. 

    \item \textbf{\texttt{Has} connections for every variable}: In this work, the interpretation of the \texttt{Has} connection is ``the (non-static) fields of a class''. However, the implementation of DP-CORE considers all variables of a class as candidates for the \texttt{has} connection, which includes local variables instantiated or used in the implementation of their methods.

    \item \textbf{``Full'' naming issues - missing connections}: The implementation of DP-CORE presented some issues with the naming of classes and objects. At first, it caches all the parsed classes in a data structure by their simple names. Then, when building the connections between classes, if faced with the full name of a class, it will look for parsed classes by their full name when they were cached with their simple name. Therefore, the object is not found, and the correct connections are not built, which leads to missing the detection of some instances. DP-LARA uses the ``full name'' in every step of the detection approach, which makes it able to detect these cases.

    \item \textbf{``Full'' naming issues - conflicts}: Another issue due to the usage of simple names of objects by DP-CORE is conflicts during the parsing phase. The parsed information is stored in a \texttt{HashMap} where the keys are the names of the objects. However, classes with equal names can be located in different packages, and inner classes in different outer classes can have equal names. With DP-CORE's usage of simple names, there could be a conflict of names -- previously parsed object would be overwritten by the last parsed object. It has two consequences: the information of the first objects would be lost, and it would result in a connection between objects that do not exist. As previously stated, DP-LARA uses the ``full name'' in every step of the detection approach, which allows it to avoid these conflicting issues.

\end{itemize}

In summary, despite the slight differences in detection results, the inspection of the differences between both tools reaffirms that the detection performance of DP-LARA is equal to DP-CORE. The sources of variability are limitations in the parsing technology used in DP-CORE (\textit{e.g.}, without ``full name''), and inaccuracies of DP-CORE's implementation in applying the UML-based concepts (\textit{e.g.}, ``static methods'' not filtered) and in handling some Java's peculiarities (\textit{e.g.}, inner and anonymous classes). Therefore, since DP-LARA is able to properly handle them, we can state that DP-LARA has a better design pattern detection performance than DP-CORE, or more specifically, DP-LARA has a better implementation of the key concepts defined by DP-CORE than their own originator.

\subsection{Detection Consistency}\label{evaluation:detection-consistency}

\begin{figure}[htbp]
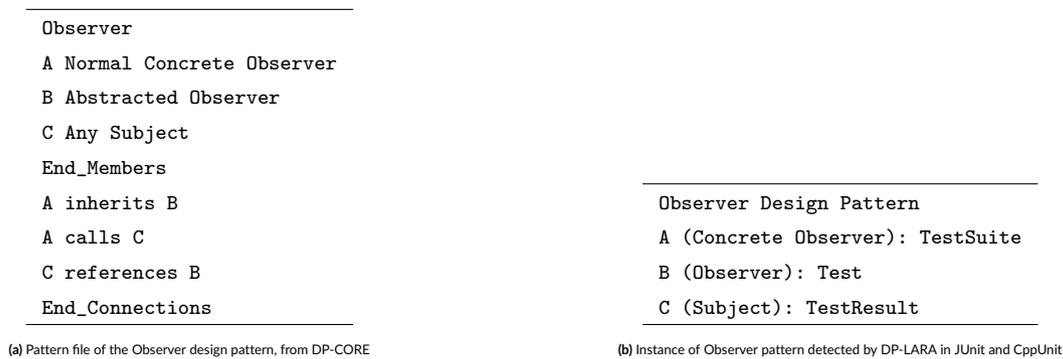

\begin{subfigure}{.48\textwidth}
\centering
\begin{tabular}{l} 
 \hline
\texttt{Observer} \\
\texttt{A Normal Concrete Observer} \\
\texttt{B Abstracted Observer} \\
\texttt{C Any Subject} \\
\texttt{End\_Members} \\
\texttt{A inherits B} \\
\texttt{A calls C} \\
\texttt{C references B} \\
\texttt{End\_Connections} \\
 \hline
\end{tabular}
\caption[Pattern file for DP-CORE of the \textsc{Observer} design pattern.]{Pattern file of the \textsc{Observer} design pattern, from DP-CORE}
\label{figure:dpcore-observer-pattern-file}
\end{subfigure}%
\begin{subfigure}{.48\textwidth}
\centering
\begin{tabular}{l} 
 \hline
 \texttt{Observer Design Pattern} \\
 \texttt{A (Concrete Observer): TestSuite} \\
 \texttt{B (Observer): Test} \\
 \texttt{C (Subject): TestResult} \\
 \hline
\end{tabular}
\caption{Instance of \textsc{Observer} pattern detected by DP-LARA in JUnit and CppUnit}
\label{figure:cppunit-observer}
\end{subfigure}

\caption[\textsc{Observer} pattern definition and instance detected by DP-LARA.]{\textsc{Observer} pattern definition and instance detected by DP-LARA.}
\label{figure:observer-pattern}
\end{figure}

The evaluation with code snippets showed that DP-LARA has equal detection range as DP-CORE in Java and C++, but experiments with larger codebases were only carried out with Java projects. In order to evaluate the consistency of detection in multiple languages, the C++ and Java projects need to be extremely similar in their design and code structure, and information regarding the use of design patterns should be available. Therefore, CppUnit and JUnit projects were selected, in particular the versions when they were the most similar. DP-LARA analyzed both projects and the results are displayed in Table~\ref{table:eval:detection-results-cppunit}. The results demonstrate that only one instance of the \textsc{Observer} design pattern is detected in both projects, and both detected the collection of classes displayed in Figure~\ref{figure:cppunit-observer}.

\begin{table*}[htbp]
\begin{center}
\caption[Number of instances detected by DP-LARA in Java and C++ projects.]{Number of Design Pattern instances detected by DP-LARA in Java and C++, for the projects JUnit 3.4 and CppUnit 1.9, respectively.}
%\small
\setlength{\tabcolsep}{6pt} % Default value: 6pt
\renewcommand{\arraystretch}{1.4} % Default value: 1
\begin{tabular}{l  c  c} 
 \hline
 & 
 \makecell{JUnit} &
 \makecell{CppUnit} \\
  % & \multicolumn{1}{c}{C} & \multicolumn{1}{c}{L} & \\ [0.5ex]
 \hline
 \textsc{Abstract Factory} & 0 & 0 \\ 
 \textsc{Builder}          & 0 & 0 \\
 \textsc{Bridge}           & 0 & 0 \\
 \textsc{Command}          & 0 & 0 \\
 \textsc{Observer}         & 1 & 1 \\
 \textsc{Visitor}          & 0 & 0 \\
 \hline
\end{tabular}
\label{table:eval:detection-results-cppunit}
\end{center}
\end{table*}

Information regarding the design patterns used in JUnit is publicly documented\footnote{\url{https://web.archive.org/web/20221128004056/https://junit.sourceforge.net/doc/cookstour/cookstour.htm}}, and, despite being based on a more recent version (v3.8) than the one used in this experiment (v3.4), we can infer about them. It does not claim to use the \textsc{Observer} pattern, even though DP-LARA (and even DP-CORE) detected instances of them. By observing its definitions (illustrated in Figure~\ref{figure:dpcore-observer-pattern-file}) and contrasting it with the definition of the remaining patterns, it only has 3 members and 3 connection types, which is the lowest number of conditions among them. This relaxed definition may lead to higher false positive detections.

\begin{figure}[htbp]
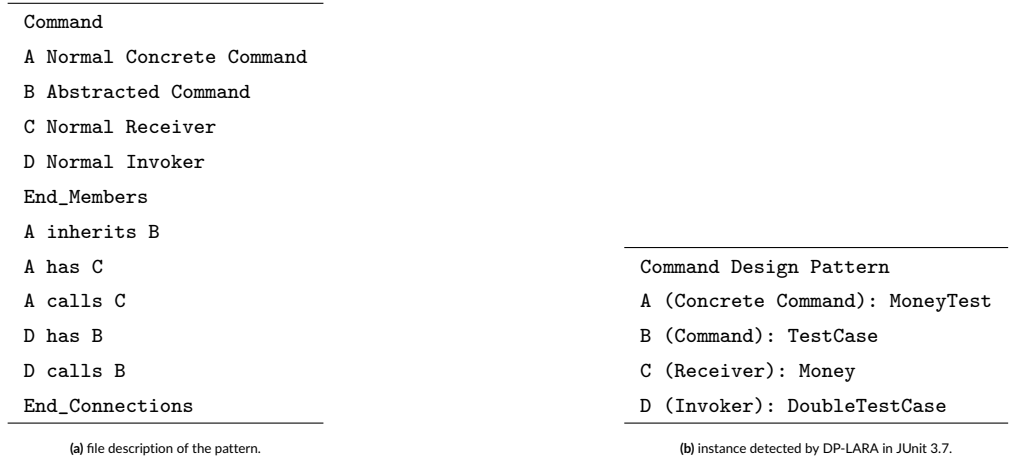

\begin{subfigure}{.48\textwidth}
\centering
%\small
\begin{tabular}{l}
 \hline
\texttt{Command} \\
\texttt{A Normal Concrete Command} \\
\texttt{B Abstracted Command} \\
\texttt{C Normal Receiver} \\
\texttt{D Normal Invoker} \\
\texttt{End\_Members} \\
\texttt{A inherits B} \\
\texttt{A has C} \\
\texttt{A calls C} \\
\texttt{D has B} \\
\texttt{D calls B} \\
\texttt{End\_Connections} \\
 \hline
\end{tabular}
\caption[*]{file description of the pattern.}
\label{figure:dpcore-command-pattern-file}
\end{subfigure}%
\begin{subfigure}{.48\textwidth}
\centering
%\small
\begin{tabular}{l} 
 \hline
 \texttt{Command Design Pattern} \\
 \texttt{A (Concrete Command): MoneyTest} \\
 \texttt{B (Command): TestCase} \\
 \texttt{C (Receiver): Money} \\
 \texttt{D (Invoker): DoubleTestCase} \\
 \hline
\end{tabular}
\caption{instance detected by DP-LARA in JUnit 3.7.}
\label{figure:cppunit-command}
\end{subfigure}

\caption[\textsc{Command} pattern definition and instance detected by DP-LARA.]{\textsc{Command} pattern definition and instance detected by DP-LARA.}
\label{figure:command-pattern}
\end{figure}

Even though DP-LARA showed consistent results in detecting DPs in multiple languages, the results were not fully encouraging, as only one (false positive) instance was detected. One of the reasons is the size of the projects that, despite being open-source and developed with intents non illustrative of design patterns, were just slightly larger than the code snippets previously used, with around ten classes. Among the DPs JUnit claims to use, as reported in the publicly available information, this work only has the definition for the \textsc{Command} pattern. Definitions for the remaining patterns could be easily added by using the UML-based representation language, but it was out of the scope of this work. Nevertheless, instances of this design pattern were not detected, which was due to the types of classes that were analyzed. Only the base classes of the framework were analyzed because in the versions selected for JUnit and CppUnit, the examples of usage were limited and different from each other. Therefore, it could not detect instances of the \textsc{Command} pattern because the projects did not have any ``Concrete Command'' (see Figure~\ref{figure:dpcore-command-pattern-file}).

Therefore, later versions of the projects were used to verify if, without restrictions in being closely implemented in both languages, DP-LARA would be able to detect instances of the \textsc{Command} pattern. In a previous experiment in Section~\ref{evaluation:detection-performance}, one of P-MARt projects is a later version of JUnit (v3.7) with examples of usage, and DP-LARA was able to detect instances of the pattern (see Figure~\ref{figure:cppunit-command}). In C++, a later version of CppUnit (v1.12) was used, which also has examples of usage. This experiment also allowed us to assess the capacity of DP-LARA to analyze larger C++ projects, since this project has more than 100 classes.

The results in Table~\ref{table:eval:detection-results-cppunit-latest} demonstrate that DP-LARA is capable of evaluating large C++ projects that have typical C++ peculiarities that the code snippets used earlier did not have: splitting the definitions and implementations respectively in \emph{header} and \emph{source} files, or the inclusion of \emph{header guards}. In addition, instances of the \textsc{Command} pattern were detected, which also corresponded to examples of usage.

\begin{table*}[htbp]
\begin{center}
\caption[Number of instances detected by DP-LARA in CppUnit]{Number of Design Pattern instances detected by DP-LARA in CppUnit 1.12.}
%\small
\setlength{\tabcolsep}{6pt} % Default value: 6pt
\renewcommand{\arraystretch}{1.4} % Default value: 1
\begin{tabular}{l  c} 
 \hline
 & \makecell{CppUnit} \\
 \hline
 \textsc{Abstract Factory} & 0 \\ 
 \textsc{Builder}          & 0 \\
 \textsc{Bridge}           & 0 \\
 \textsc{Command}          & 2 \\
 \textsc{Observer}         & 1 \\
 \textsc{Visitor}          & 0 \\
 \hline
\end{tabular}
\label{table:eval:detection-results-cppunit-latest}
\end{center}
\end{table*}

\subsection{Multi-Language Extensibility}\label{evaluation:extensibility}

Finally, the capacity of DP-LARA to add support to new programming languages was assessed by comparing it against other multi-language detection approaches found in the literature. In Related Work (Section~\ref{related-work}), six different tools were identified as having multi-language capabilities. Even though not all works were fully materialized to support more than one language (for convenience reasons or for not being a property they were purposely seeking), they followed a common multi-layered architecture. 

This evaluation study is divided into two phases. Firstly, the necessary steps to extend DP-LARA are described so that the effort required can be understood. Then, the effort to extend the approaches found in the literature is inferred from their publications and other public documentation, and compared with DP-LARA. The comparative analysis is supported by qualitative data about the number of tasks required to produce the abstraction representation of the software system, and how complex and different the representation is from the starting representation, which often is the AST of the software system.

\subsubsection{Extensibility of DP-LARA}\label{chap:evaluation:extensibility:lara}

Recalling the architecture of DP-LARA, the extraction of the information from the source code and the detection algorithm are executed on LARA with the LARA Common Language.
LARA acts as an abstraction of the programming languages by providing a virtual AST, whose specification is based on OOP concepts that are commonly shared among languages. The abstraction step succinctly consists of mapping the source code to this virtual AST --- the key location when extending to new languages.

We extended the common specification with attributes and join points that were necessary for the detection of design pattern instances.

\begin{figure*}[htbp]
% \centerline{\includegraphics[width=0.8\linewidth]{images/LARA-Diagrams-Pipeline-Extension-Clava.jpg}}
\centerline{\includegraphics[width=0.8\linewidth]{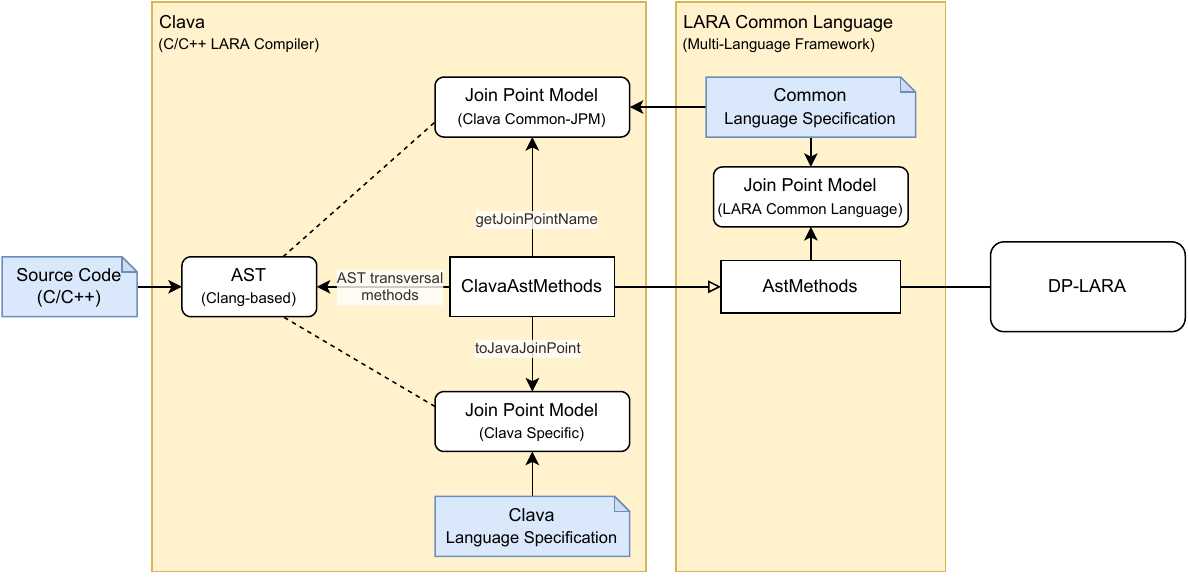}}
\caption{LARA Compiler extension and compliant with LARA Common Language.}
\label{figure:clava_lara-common-model}
\end{figure*}

LARA already supported several programming languages, in particular, compilers for C/C++ (Clava) and Java (Kadabra) were already consolidated\added[id=HA]{, but also JavaScript and Matlab. They have different states of compliance to the LARA Common Language, as they follow independent developments and have different applications}. Each compiler has a specific language specification, therefore, when providing implementation to the join point model of the common language, much of it could be reused, with the helper method \texttt{toJavaJoinPoint} (see Figure~\ref{figure:clava_lara-common-model}). It converts the AST node into the join point of the original language specification, and thus allowing its reuse. In addition, LARA has generators which create template join points for the common language specification, therefore the developer is only required to provide implementation for the attributes of interest, and does not need to worry about the overall structure of the class join point.

With this framework structure, code generators, and helper methods, the implementation effort lies solely on mapping the AST nodes of the specific LARA compiler with the names of the corresponding join points and reusing the implementation of a reduced number of key attributes. One could say that the learning effort is actually bigger than the coding effort, but since LARA is an aspect-oriented framework where compilers can be seen as providers of an abstraction layer over the target languages (\textit{e.g.} C++, Java),  %extensions of their target languages, %the learning curve would still be lower. It would depend on the experience of the developers and their familiarity with the languages.
we considered that the learning curve will be lower for developers who are familiar with the target language and corresponding ASTs.

\begin{figure*}[htbp]
% \centerline{\includegraphics[width=0.8\linewidth]{images/LARA-Diagrams-Pipeline-Extension-New.jpg}}
\centerline{\includegraphics[width=0.8\linewidth]{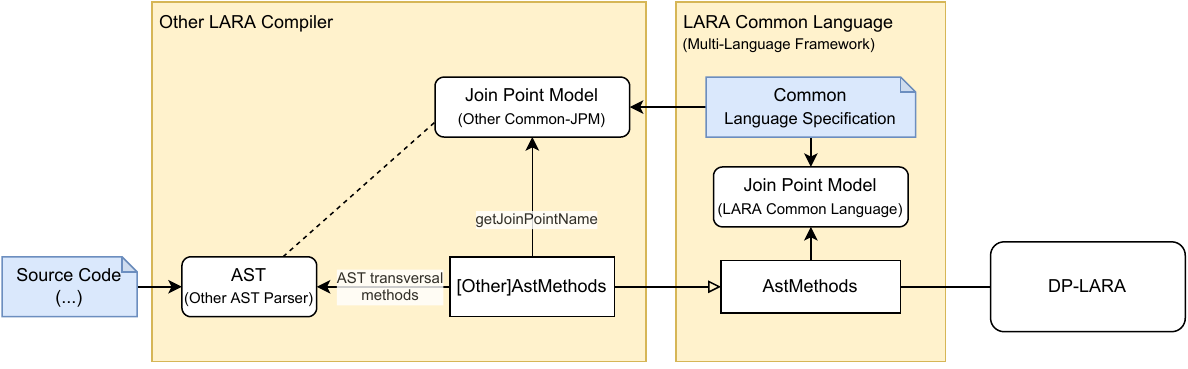}}
\caption{New LARA Compiler, compliant with LARA Common Language.}
\label{figure:lara-compiler-support}
\end{figure*}

Nevertheless, there are languages that LARA does not have support for (e.g C\#). In order to add them for the sole purpose of multi-language DPD, the absence of helper methods (since no specific LARA compiler would exist) would require the implementation of a language-specific compiler (see Figure~\ref{figure:lara-compiler-support}).

It could use the common specification directly. Therefore, the effort of extending DP-LARA to a new language unsupported by LARA would be in adding a compiler to LARA, selecting a parser that can output a Java-based AST, mapping the AST nodes to the common language join points, and providing implementation to the key attributes. Apart from the effort of selecting and learning to use an AST parsing tool, which is common to all multi-language approaches, the difference lies in the mapping of the AST nodes and in providing the missing implementation. In simple terms, this effort can be viewed as mapping the tools' AST to the LARA virtual AST; therefore, the mapping is very straightforward, and the implementations are short since the required attributes usually have equivalents in the AST tool or are not very difficult to obtain.

Finally, the LARA Compiler of the new language would need to have its own implementation of the interface \texttt{AstMethods}, which is where the communication between the LARA Common Language and the compiler is handled. This is where the mapping method between AST nodes and join points would be placed. In addition, this class would also have the methods related with navigation through the AST (\texttt{getChildren}, \texttt{getRoot}, \texttt{getParent}, \texttt{getDescendants}), which would also be straightforward. 

In summary, whether the new language is already supported in the LARA framework or not, the effort of extending DP-LARA is very localized and requires reasonable coding effort. The more arduous effort is learning how the AST tool and the LARA framework work, but this is not a specific endeavor to DP-LARA. As described in the following section, multi-language approaches share this learning effort, but LARA eases it out due to its publicly available documentation and straightforward effort between the language-specific AST representation and the common join point model.

\subsubsection{Extensibility Comparison}\label{chap:evaluation:extensibility:comparison}

We were not able to obtain the code-bases of the projects of the multi-language detection tools identified in the Related Work. Therefore, the comparative analysis will be narrowed to their respective publications and other publicly available documentation, and the effort required to extend them to new languages is inferred from this information.

\begin{figure}[htbp]
\centerline{\includegraphics[width=0.55\linewidth]{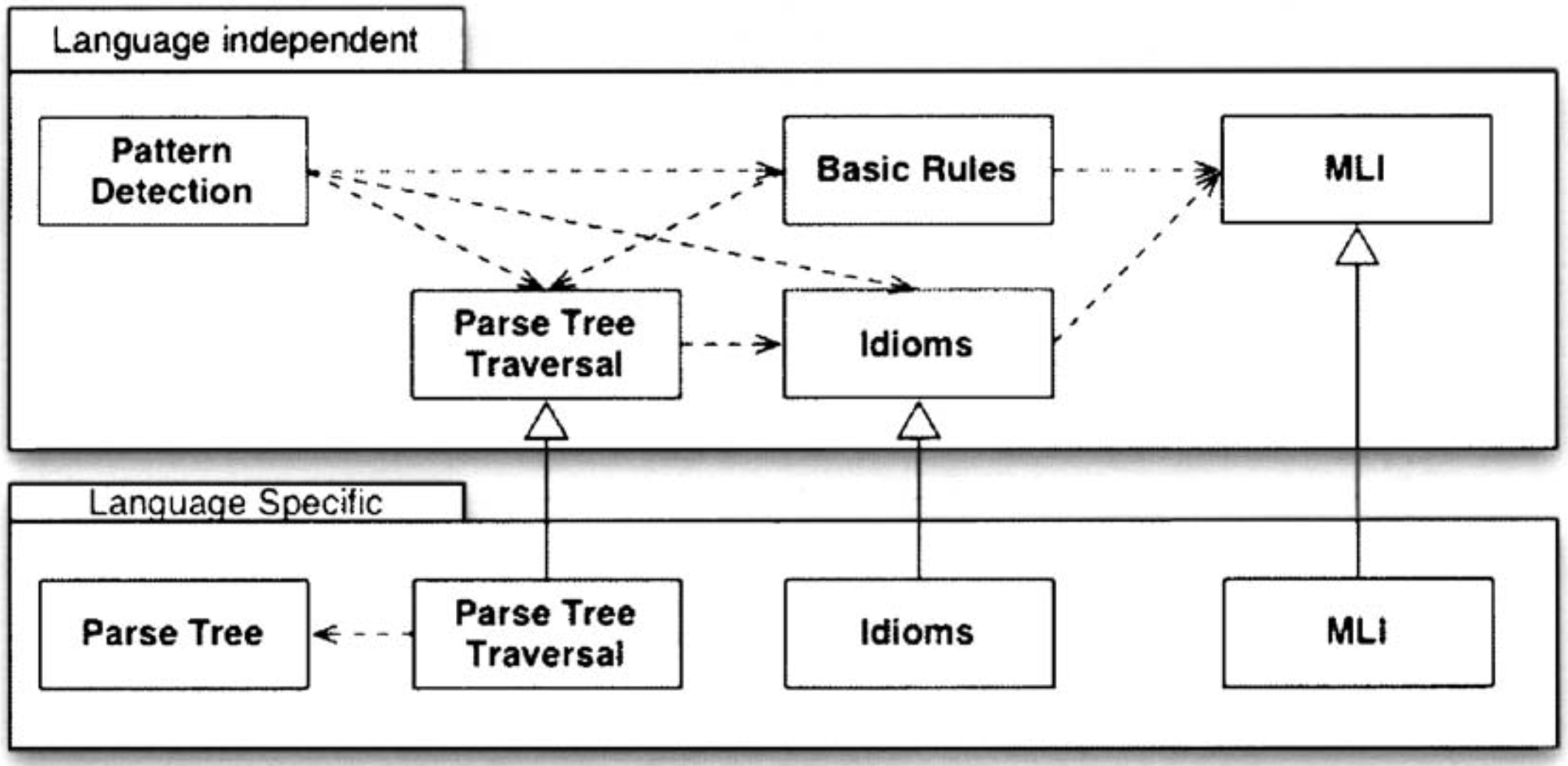}}
\caption[Architecture of SOULJava.]{Architecture of SOULJava~\cite{SOUL2002}.}
\label{figure:ml-souljava}
\end{figure}

One approach for language-independent detection is the use of a metaprogramming language, and the detection algorithm would be reasoned at this level. Fabry \& Mens~\cite{FABRY200421} used SOUL as the meta programming language, which already supported Smalltalk, and the detection algorithm consists of a series of queries and logic facts at the meta level. In their work, they extended the SOUL to support Java (SoulJAVA), and described the necessary steps. They claim there are three locations that require language-specific extensions, as illustrated in Figure~\ref{figure:ml-souljava}. The \textit{meta-level interface} provides a representational mapping between the logic meta language and the specific language, which contains 47 methods for Smalltalk and 58 for Java. The \textit{parse tree traversal} is used to build the logic parse tree at the meta level, which is typically obtained with predicates for each kind of parse tree node: 14 rules for Smalltalk tree traversal and 37 for Java traversal. Finally, the \textit{idioms} cover syntactic language differences and naming and coding conventions: 6 idioms for Smalltalk and 8 for Java. 

Even though the approaches are quite different (DP-LARA is UML-based on a virtual-AST, and SOULJava is query-based on a logic meta-programming language), there are some similarities when it comes to extending them to new languages. Both require source code parsers and mapping between the language and the meta levels. DP-LARA requires compliance with a common specification of 15 join points with 21 attributes in total, and the interface between LARA and the base language is minimal: implementation of 5 methods of the \texttt{AstMethods}, which includes 1 mapping method, and 4 common AST navigation methods. SoulJAVA required the implementation of 58 methods for the \textit{meta-level interface} whose size and effort is not disclaimed. SoulJAVA also requires implementation for the \textit{parse tree traversal} module, but they claim it can be automated to a large extent. Finally, the module \textit{idioms} also needs to have an implementation in the base language, but it would not be fairly compared with DP-LARA since it adds a dimension that was not considered in DP-LARA: syntactic analysis.

\begin{figure}[bp]
\centerline{\includegraphics[width=0.75\linewidth]{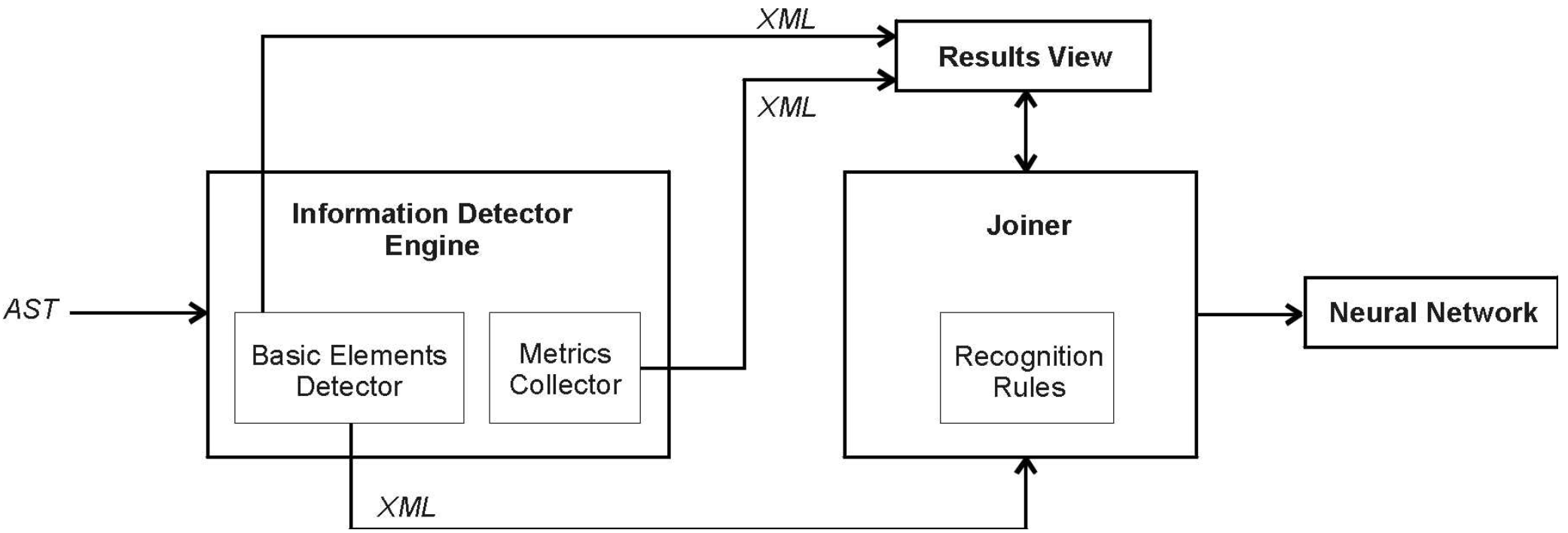}}
\caption[Architecture of nMARPLE.]{Architecture of nMARPLE~\cite{MARPLE-NET}.}
\label{figure:ml-nmarple}
\end{figure}

Arcelli \textit{et al.} developed nMARPLE~\cite{MARPLE-NET}, which follows a multi-layered approach for design pattern detection (as illustrated in Figure~\ref{figure:ml-nmarple}), which sequentially detects \textit{basic elements} from the source code, and then design pattern instances from \textit{basic elements} through graph-matching and machine learning. The modules were not fully integrated in a single application. However, the first module should output an XML file with the extracted basic elements, and this is the multi-language injection location: the \textit{basic elements detector}. The languages that implement this module should be able to detect the 16 types of basic elements, which basically are formalization of structures such as inheritance, object creation, and method invocation. Once again, we were unable to obtain the code base, but the effort could be estimated by the tools used in nMARPLE (and predecessors) and what would be required to detect the basic elements. Similar to DP-LARA, AST parsing tools were used. However, the complexity of the basic elements varies: from the basic identification of a class is abstract and its inheritance relationships to the complex differentiation of the type of method call (\textit{e.g.}, recursion, redirect, or delegate). In summary, the language-specific implementation of nMARPLE requires some level of detection algorithms, although more basic, while LARA simply requires mapping language-specific ASTs to the common AST concepts -- LARA Common Language. One could infer that the overall effort of extending nMARPLE is higher, and probably the amount of code required also.

\begin{figure}[htbp]
\centerline{\includegraphics[width=0.5\linewidth]{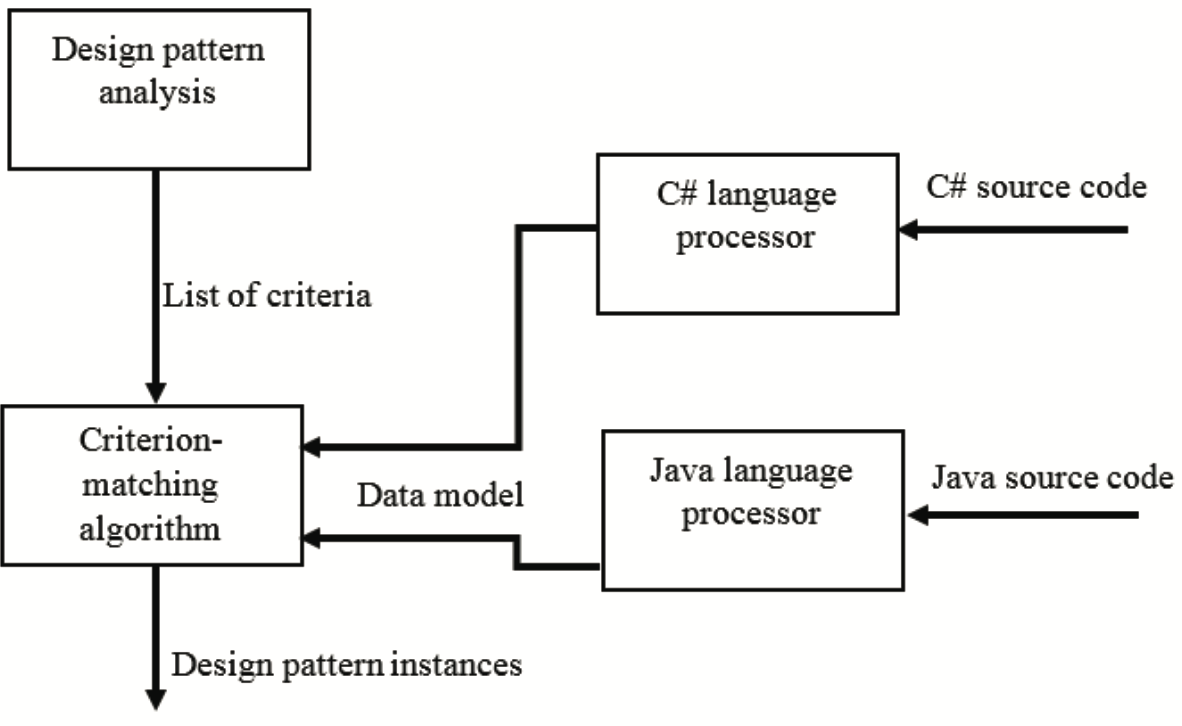}}
\caption[Architecture of ``language-neutral framework''.]{Architecture of ``language-neutral framework''~\cite{NEUTRAL}.}
\label{figure:ml-neutral}
\end{figure}

Nagy \& Kovari~\cite{NEUTRAL}, developed a \enquote{language neutral design pattern recognition tool} with some similarities to DP-LARA, as illustrated in Figure~\ref{figure:ml-neutral}. The detection algorithm uses an instance of an object-oriented data model with the necessary information for the detection algorithm. The abstraction step is done by the \textit{language processor}, which is responsible for analyzing the source code and producing an instance of the data model. From the brief description and examples provided, the data model seems to include concepts very similar to the attributes of the LARA common specification, and the language processor depends exclusively on a parsing tool (.NET Compiler Platform). However, no more details were provided and the project code was not disclosed.

CrocoPat~\cite{CrocoPat}, developed by Beyer \textit{et al.}, creates an abstract representation of the source code. This meta model consists of 4 entities (packages, methods, attributes, and packages) with a small set of attributes, and a relationship between these entities%, as illustrated in Figure~\ref{figure:ml-crocopat}
. Then, their approach is able to extract binary relations and use them in the graph-based detection algorithm. This approach has many similarities with DP-LARA. Pattern definitions use a resembling representation as DP-LARA, and the relations use similar concepts such as \textit{inherits}, \textit{calls}, or \textit{contains}. However, due to how CrocoPat is structured, the module that analyzes the source code needs to result in a relations file, which means that when extending to new languages, the module that would be re-implemented would need to not only create the meta-model but also add logic to transform it to binary decisions. In contrast, DP-LARA does a similar reasoning, when extracting the connection types, but at the abstract level, and is reused between languages. By having these fewer steps, one could infer that the effort of extending DP-LARA is reduced when compared with CrocoPat.

DeMIMA~\cite{DEMIMA}, developed by Guéhéneuc \textit{et al.} defined a domain-specific language inspired by UML, which is used to express the meta-model of the systems. First, the tool constructs an instance of the meta-model which contains object-oriented concepts from Java. Then, the metamodel is enriched with binary class relationships. In order to extend DeMIMA to new languages, language-specific AST parsing should reproduce the UML-inspired meta-model, which requires replicating this logic for every language, while LARA performs a similar logic (extraction of UML-based information) but only once at the abstract level.

\begin{figure}[htbp]
\centerline{\includegraphics[width=0.9\linewidth]{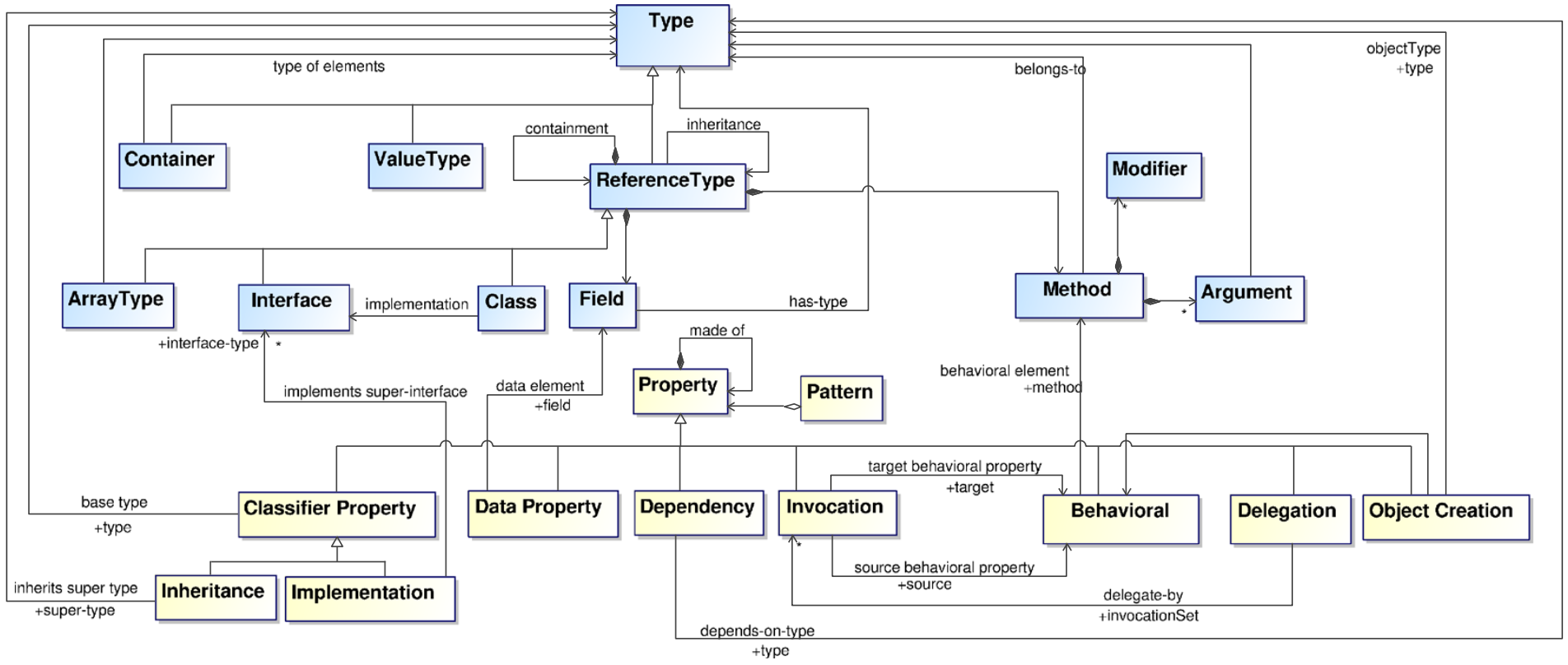}}
\caption[Meta-model of DPF.]{Meta-model of DPF~\cite{Graph-DSL}.}
\label{figure:ml-dpf}
\end{figure}

Finally, DPF~\cite{Graph-DSL} also uses a meta-model for the abstraction of the software system, as illustrated in Figure~\ref{figure:ml-dpf}. Both the meta-model and the descriptions of design patterns are modeled on a domain-specific language, and then the detection algorithm uses graph matching. The meta-model is more complex than the one from CrocoPat, as it describes the system in terms of relationships, including structural (\textit{e.g.}, inheritance, implementation, visibility) and behavioral (\textit{e.g.}, delegation and objection creation) ones. The abstraction step consists of generating an instance of the system model through AST transversal. In order to extend DPF to new languages, the abstraction step needs to be re-implemented, and, similarly to CrocoPat, this module requires logic to transform the AST to the meta-model. Once again, LARA, by having the abstraction representation (a virtual AST) closer to the language-specific representation (an AST), provides a more straightforward approach, and with significantly less effort, to the issue of extending it to new programming languages.

In summary, DP-LARA is a multi-language approach for design pattern detection that currently supports Java and C++, and it inherits the rules-satisfaction methodology (based only on static analysis) and extensibility properties to define new patterns from DP-CORE and benefits from the multi-language properties of LARA to easily extend to other programming languages. We believe the effort to extend to new languages takes less effort than other multi-language detection approaches. As summarized in Table~\ref{table:multi-language-effort-evaluation}, the abstract representation is similar to the initial representation, and the critical tasks to extend are more straightforward since most of them are simply the mapping of concepts from the language-specific AST to the abstraction of the AST provided by LARA. All the logic related to the detection of patterns is reused between different languages.

\begin{table*}[htbp]
\begin{center}
\fontsize{8}{10}\selectfont
\caption{Effort to extend approach to new specific languages}
%\small
\setlength{\tabcolsep}{6pt} % Default value: 6pt
\renewcommand{\arraystretch}{2} % Default value: 1
\begin{tabular}{l | c | l }%| l} 
 \hline 
 \textbf{Approach} & \textbf{Abstract Representation} & \textbf{Tasks} \\ [0.5ex] 
 \hline
 DP-LARA
 & \makecell[c]{Abstraction of the AST \\ (15 join points + 21 attributes) }
 % & AST Parser
 & \makecell[l]{- Mapping of the AST nodes to join points\\(and provide implementation of the attributes)\\
 - Navigation of the AST tool from the abstract level} \\ 
 \hline
 SoulJAVA 
 & Meta programming language
 % & AST Parser
 & \makecell[l]{- Mapping between logic meta language and specific\\language (58 methods) \\
  - AST transversal to build logic parse tree (37 rules)}
  \\ 
 \hline
 nMARPLE 
 & Collection of basic elements
 % & AST Parser
 & \makecell[l]{- AST Transversal\\
 - Detection of basic elements (16 types)}
  \\ 
 \hline
 Nagy \& Kovari 
 & \makecell[c]{Data Model\\(object oriented concepts)}
 % & AST Parser
 & \makecell[l]{- AST transversal to produce instance of\\data model}
  \\ 
 \hline
 CrocoPat
 & \makecell[c]{Meta Model\\(4 entities, small set\\of attributes and relationships)}
 % & AST Parser
 & \makecell[l]{- AST transversal to generate meta model \\
 - Extract binary relations}
  \\ 
 \hline
 DeMIMA
 & \makecell[c]{Meta Model\\(UML-based, complex)}
 % & AST Parser
 & \makecell[l]{- AST transversal to generate meta model\\
 - Extract binary class relationships}
  \\ 
 \hline
 DPF
 & \makecell[c]{Meta Model\\(Graph-based and complex)}
 % & AST Parser
 & \makecell[l]{- AST transversal to generate meta model }
  \\ 
 \hline
\end{tabular}
\label{table:multi-language-effort-evaluation}
\end{center}
\end{table*}

\subsection{Threats to Validity}

Each experiment conducted in the evaluation of the approach seeked optimal experimental conditions so that results could have a strong validation. The threats to validity were identified and efforts to mitigate them were made.

\begin{itemize} 
\setlength\itemsep{0em}

    \item \textbf{Projects Selection}: The approach was evaluated against the DP-CORE -- the selected design pattern detection approach to which multi-language support was added. The projects chosen were from DP-CORE's publication and from the repository of pattern-like micro-structures, P-MARt. The former projects allowed one to evaluate the approach as closely as possible to the conditions of the publication. The latter allowed one to evaluate both DP-CORE and DP-LARA with a baseline that many publications in this field use. In addition, these projects are open-source, and their size varies; thus, they provide a good balance between mimicking industry projects and being reproducible.
    
    For the evaluation of the capability of the approach to detect design pattern instances, we also used code snippets. For Java, snippets made available from DP-CORE's project were used so that the evaluation could also be compared against DP-CORE. The snippets in Java could be ported to C++, but bias could be introduced in the process which would be a threat to validity. Thus, we extracted equivalent code snippets from \textit{Refactoring.guru}, an informative website that addresses the topic of design patterns.
    
    The same issue arose when evaluating the consistency of the approach in Java and C++. The ideal conditions were to select projects with equal implementations in both languages. To manually port a project from one language to the other would introduce bias that would threaten the validity of the results. Thus, the option was to select open-source projects that were ported from Java to C++ or vice-versa, and therefore, JUnit and CppUnit satisfied these conditions. After inspection of the code, the latest versions had diverged, which indicates that after porting they followed independent developments. Therefore, the versions closest to the date of the port were selected. Even though they have a fraction of the size of the latest versions, they have very similar structures.
    
    \item \textbf{DP-CORE version}: DP-CORE satisfied several conditions defined at the start of this work: structured methodology where multi-language support could be easily injected, flexibility to define new definitions of design patterns, and is publicly available. However, the latest available version could be different from what was used in their publication. Even though the DP-CORE detection results in this work were very close to what was reported, the slight differences might be caused by either this difference of versions or by the differences of the versions of the projects analyzed. In order to mitigate these threats, the comparison of DP-LARA was done against the latest version of DP-CORE by analyzing the extracted projects, and not against what was published in their article.
    
    \item \textbf{Comparison with Multi-Language Approaches}: The multi-language approaches to design pattern detection were selected after the literature review, and even though it was an extensive and methodical work, it can not be affirmed that no approach was missed. On the other hand, the projects of the selected approaches were not obtained, therefore the comparison against DP-LARA was limited to what was reported. The approaches detailed differently their methodology to provide multi-language support, and each used different technologies. Therefore, the comparison done in this work excluded the technicalities that each AST parser presented and focused solely on the effort required to transform an AST representation into an abstract representation. Nevertheless, the detection performance sought by the authors might influence the complexity of the abstract representation, which was a factor that could be more properly addressed if inspecting the respective code base was possible and evaluating them in the same conditions as DP-LARA.

\end{itemize}

\section{Discussion}\label{discussion}

DP-LARA consisted of injecting multi-language support to an established design pattern detection approach, DP-CORE, and experiments demonstrated that both have similar detection results. After inspection, the sporadic differences were due to limitations of the parsing technology and inaccuracies in the implementation that did not account for all the language peculiarities (\textit{e.g.}, anonymous classes, static methods, and fields), which are justifiable causes when implementations are ported.

Moreover, experiments successfully demonstrated the consistency of DP-LARA in detecting design patterns in Java and C++ source code. At a small scale, DP-LARA had the exact same results for equivalent code snippets - it was able to detect all instances of the supported design patterns in their respective code snippets, and even detected the same ``false positive'' results as DP-CORE in both languages. Scaling this experiment to open-source projects, closely similar projects were selected (CppUnit and JUnit), and once again, DP-LARA was able to detect the exact same instances in both projects.

DP-CORE satisfied the criteria set for the selection of a detection algorithm, to whom this work would add support for multi-language. Nonetheless, one of the main reasons for its selection was that it is publicly available, which eased the analysis, code inspection, and development phases of this project. It was also paramount in the evaluation phase because the experiments could be executed with DP-LARA and DP-CORE under the exact same conditions. Despite the selection of DP-CORE, LARA can be easily extended to other detection approaches found in the literature. By providing a virtual AST (specified as join points and attributes in the common specification) that can be extended with low effort, and since most approaches use AST parsing somewhere in their methodology, a similar procedure can be used to inject LARA: detect the injection point in the tool, extend LARA Common Specification accordingly, and replace it with an aspect in the LARA Common Language.

In fact, other detection methodologies possess multi-language properties. Despite not being able to inspect their projects, the comparative evaluation between LARA and their publications allowed one to make assumptions about the effort required for the extension of the approaches to new languages. LARA, by having the abstract representation closer to the source code than other multi-language approaches, decreases the effort of extending to new programming languages. LARA provides a virtual AST whose much of the effort is mapping the nodes of the AST parsing tools with equivalent join points in the common specification. Additionally, much of the implementation of the attribute of the common model is already given by methods of the equivalent AST nodes; the effort in these cases is simply differences in nomenclature or output format. On the other hand, other approaches have more complex abstract representations (\textit{e.g.}, data models or logic languages) that require more steps and more complex logic than mapping equivalent concepts.

Finally, since the data models of other approaches are inspired by OO concepts, they could be instanced using the virtual AST provided by the LARA Common Language. In their original publications, language-specific ASTs are used, but the required information could be extracted using LARA. In other words, the logic for the creation of the data models could be moved to a single point in a higher abstraction level (using LARA), instead of replicating it for each programming language.

\section{Conclusion}\label{conclusion}

% KEY BENEFITS
In the software development lifecycle, automatic tools for the detection of DPs can be significantly useful in code comprehension. While the complexity of projects has increased, often encompassing multiple languages, most DPD tools are specific to a single language. In addition, the tools very seldomly are available, and the few multi-language approaches developed often require significant effort to extend to a new language, which hinders their evolution.

DP-LARA is a DPD tool which benefits users and developers. DP-LARA allows users to use a single tool for multiple programming languages. In addition, the effort of extending DP-LARA to new languages is reduced due to having the abstract representation closer to the source code than other multi-language approaches. The effort required is limited to mapping the language-specific AST to the virtual AST of OO concepts, specified in the LARA Common Language.

%
% CHALLENGES
Despite having similar detection performances as its language-specific predecessor, DP-LARA has some challenges in performance. During the extraction of information, the context changes in the communication between the LARA Common Language and the language-specific LARA Compilers are computationally expensive. In addition, the main logic of the detection, including the recursive detection algorithm, is described in JavaScript, which currently runs on top of an interpreter written in Java (i.e., GraalVM), which can be a bottleneck.

%
% FUTURE WORK
There are several future steps that could improve this multi-language code analysis tool for the detection of design pattern instances:

\begin{itemize} 
\setlength\itemsep{0em}
    
    \item \textbf{Support for other languages} --- Apart from Java and C/C++, LARA has support for other programming languages; thus, DP-LARA could be extended either by making the remaining LARA Compilers compliant with the updated LARA Common Specification or by developing new LARA Compilers with the exclusive purpose of multi-language code analysis.

    \item \textbf{Improvement of Detection Algorithm} --- DP-CORE has the flexibility to add new or update existing definitions of DPs, aiming for better detection results. In addition, the detection algorithm itself could be improved by, for instance, extending the representation language and extractable information defined by DP-CORE (\textit{e.g.}, static fields/methods, arrays/collections). Dynamic analysis could also be added in order to extract behavioral information, which would imply multi-language instrumentation of the source code, a property that LARA already supports.
    
    \item \textbf{Improvement of Computational Performance} --- Despite not being statistically assessed by us, DP-LARA has longer execution times than DP-CORE, sometimes by an order of magnitude. This work contributes a first implementation, which does not focus on performance; several points that can be addressed, such as algorithmic improvement, optimization of the framework, or changing the JavaScript interpreter.
    
\end{itemize}

Finally, the evaluation methodology could be improved either by comparing it with other detection tools or by conducting case studies in a real-world context.

\begin{itemize} 
\setlength\itemsep{0em}
    
    \item \textbf{Comparison with other detection tools} --- Analyzing and comparing the detection performance of DP-LARA against other detection tools could serve as the baseline for future work. DP-LARA and other tools would need to analyze a repository of pattern-like micro-structures manually collected and classified by experts for each language they support. The repository P-MARt was used, but it only contains Java projects, therefore other labeled public repositories would be important.
    
    \item \textbf{Case Study on the adoption and maintenance of the multi-language code analysis tool} --- Running a case study in a real-world context would contribute to validating that a multi-language LARA-based code analysis tool would reduce the effort of supporting new languages while promoting consistent results across the different languages it supports. The feasibility of adopting a single detection tool for all projects within a company, and the effort of developers to maintain DP-LARA would be investigated.
    
\end{itemize}

\bibliography{references}

\begin{thebibliography}{10}
\providecommand \doibase [0]{http://dx.doi.org/}%

\bibitem{zhang2012effectiveness}
Zhang C, Budgen D. What Do We Know about the Effectiveness of Software Design
  Patterns?. {\it IEEE Transactions on Software Engineering} 2012\string;
  38(5)\string: 1213-1231.
\newblock \href {\doibase 10.1109/TSE.2011.79} {doi: 10.1109/TSE.2011.79}

\bibitem{lemos2024live}
Lemos F, Correia FF, Aguiar A, Queiroz PG. Live software documentation of
  design pattern instances. {\it PeerJ Computer Science} 2024\string;
  10\string: e2090.

\bibitem{LARA}
Pinto P, Carvalho T, Bispo J, Ramalho MA, Cardoso JM. Aspect composition for
  multiple target languages using {LARA}. {\it Computer Languages, Systems \&
  Structures} 2018\string; 53\string: 1-26.
\newblock \href {\doibase https://doi.org/10.1016/j.cl.2017.12.003} {doi:
  https://doi.org/10.1016/j.cl.2017.12.003}

\bibitem{DPDReview}
Yarahmadi H, Hasheminejad SMH. Design pattern detection approaches: a
  systematic review of the literature. {\it Artificial Intelligence Review}
  2020\string; 53.
\newblock \href {\doibase 10.1007/s10462-020-09834-5} {doi:
  10.1007/s10462-020-09834-5}

\bibitem{DPCORE}
Diamantopoulos T, Noutsos A, Symeonidis A. DP-CORE: A Design Pattern Detection
  Tool for Code Reuse. In: ; 2016

\bibitem{PINOT}
Shi N, Olsson R. Reverse Engineering of Design Patterns from Java Source Code.
  In: ; 2006\string: 123-134

\bibitem{FUJABA}
Niere J, Wadsack J, Wendehals L. Handling large search space in pattern-based
  reverse engineering. In: ; 2003\string: 274-279

\bibitem{Automatic-AST}
Heuzeroth D, Holl T, Hogstrom G, Lowe W. Automatic design pattern detection.
  In: ; 2003\string: 94-103

\bibitem{pree1995design}
Pree W. {\it Design patterns for object-oriented software development}.
\newblock ACM Press/Addison-Wesley Publishing Co. .
\newblock 1995.

\bibitem{JFREEDOM}
Flores N, Aguiar A. JFREEDOM: a Reverse Engineering Tool to Recover Framework
  Design. {\it Proceedings of the 6th European Conference on Object-Oriented
  Programming} 2005.

\bibitem{MetaPatterns}
Hayashi S, Katada J, Sakamoto R, Kobayashi T, Saeki M. Design Pattern Detection
  by Using Meta Patterns. {\it IEICE Transactions} 2008\string; 91-D\string:
  933-944.
\newblock \href {\doibase 10.1093/ietisy/e91-d.4.933} {doi:
  10.1093/ietisy/e91-d.4.933}

\bibitem{FeatureMaps}
Thaller H, Linsbauer L, Egyed A. Feature Maps: A Comprehensible Software
  Representation for Design Pattern Detection. In: ; 2019\string: 207-217

\bibitem{Feature-based}
Nazar N, Aleti A, Zheng Y. Feature-based software design pattern detection.
  {\it Journal of Systems and Software} 2022\string; 185\string: 111179.
\newblock \href {\doibase https://doi.org/10.1016/j.jss.2021.111179} {doi:
  https://doi.org/10.1016/j.jss.2021.111179}

\bibitem{ML-Classification}
Chihada A, Jalili S, Hasheminejad SMH, Zangooei MH. Source code and design
  conformance, design pattern detection from source code by classification
  approach. {\it Applied Soft Computing} 2015\string; 26\string: 357-367.
\newblock \href {\doibase https://doi.org/10.1016/j.asoc.2014.10.027} {doi:
  https://doi.org/10.1016/j.asoc.2014.10.027}

\bibitem{ML-TreeBaseAlgorithm}
Mhawish M, Gupta M. Software Metrics and tree-based machine learning algorithms
  for distinguishing and detecting similar structure design patterns. {\it SN
  Applied Sciences} 2020\string; 2\string: 11.
\newblock \href {\doibase 10.1007/s42452-019-1815-3} {doi:
  10.1007/s42452-019-1815-3}

\bibitem{Graph-InexactMatching}
Gupta M, Rao RS, Tripathi AK. Design Pattern Detection using inexact graph
  matching. In: ; 2010\string: 211-217.

\bibitem{Graph-Greedy}
Gupta M. Design pattern mining using greedy algorithm for multi-labelled
  graphs. {\it IJICT} 2011\string; 3\string: 314-323.
\newblock \href {\doibase 10.1504/IJICT.2011.043627} {doi:
  10.1504/IJICT.2011.043627}

\bibitem{Graph-CrossCorrelation}
Gupta M, Pande A, Rao RS, Tripathi A. Design Pattern Detection by normalized
  cross correlation. In: ; 2010\string: 81-84

\bibitem{Graph-DPDetect}
Singh J, Gupta M. Design Pattern Detection Using Dpdetect Algorithm. {\it
  International Journal of Innovative Technology and Exploring Engineering}
  2019.

\bibitem{Graph-Hybrid}
Singh J, Chowdhuri SR, Bethany G, Gupta M. Detecting design patterns: a hybrid
  approach based on graph matching and static analysis. {\it Information
  Technology and Management} 2021.
\newblock \href {\doibase 10.1007/s10799-021-00339-3} {doi:
  10.1007/s10799-021-00339-3}

\bibitem{Graph-Similarity}
Tsantalis N, Chatzigeorgiou A, Stephanides G, Halkidis ST. Design Pattern
  Detection Using Similarity Scoring. {\it IEEE Transactions on Software
  Engineering} 2006\string; 32(11)\string: 896-909.
\newblock \href {\doibase 10.1109/TSE.2006.112} {doi: 10.1109/TSE.2006.112}

\bibitem{Graph-TemplateMatching}
Dong J, Sun Y. Design Pattern Detection by Template Matching. In: ;
  2008\string: 765-769

\bibitem{MARPLE}
Arcelli F, Cristina L. Enhancing Software Evolution through Design Pattern
  Detection. In: ; 2007\string: 7-14

\bibitem{MARPLE-NET}
Arcelli F, Franzosi D, Raibulet C. .NET Reverse Engineering with MARPLE. In: ;
  2010\string: 227-231

\bibitem{FABRY200421}
Fabry J, Mens T. Language-independent detection of object-oriented design
  patterns. {\it Computer Languages, Systems \& Structures} 2004\string;
  30(1)\string: 21-33.
\newblock Smalltalk Language\href {\doibase
  https://doi.org/10.1016/j.cl.2003.09.002} {doi:
  https://doi.org/10.1016/j.cl.2003.09.002}

\bibitem{NEUTRAL}
Nagy A, Kovari B. Programming language neutral design pattern detection. In: ;
  2015\string: 215-219

\bibitem{CrocoPat}
Beyer D, Lewerentz C. {\it CrocoPat: A Tool for Efficient Pattern Recognistion
  in Large Object Oriented Programs}.
\newblock Citeseer .
\newblock 2003.

\bibitem{Graph-DSL}
Bernardi M, Cimitile M, Di~Lucca G. Design Patterns Detection Using a
  DSL-driven Graph Matching Approach. {\it Journal of Software: Evolution and
  Process} 2014\string; in press.
\newblock \href {\doibase 10.1002/smr.1674} {doi: 10.1002/smr.1674}

\bibitem{bispo2020clava}
Bispo J, Cardoso JM. Clava: C/C++ source-to-source compilation using LARA. {\it
  SoftwareX} 2020\string; 12\string: 100565.

\bibitem{carvalho2023dsl}
Carvalho T, Bispo J, Pinto P, Cardoso JM. A DSL-based runtime adaptivity
  framework for Java. {\it SoftwareX} 2023\string; 23\string: 101496.

\bibitem{reis2020compilation}
Reis L, Bispo J, Cardoso JM. Compilation of MATLAB computations to CPU/GPU via
  C/OpenCL generation. {\it Concurrency and Computation: Practice and
  Experience} 2020\string; 32(22)\string: e5854.

\bibitem{LARAM}
Teixeira G, Bispo J, Correia FF. Multi-Language Static Code Analysis on the
  {LARA} Framework. In: SOAP 2021. Association for Computing Machinery; 2021;
  New York, NY, USA\string: 31–36.

\bibitem{P-MARt}
Guéhéneuc YG. P-MARt: Pattern-like Micro Architecture Repository. {\it 1st
  EuroPLoP Focus Group on Pattern Repositories} 2007.

\bibitem{SOUL2002}
Wuyts R. {\it A logic meta-programming approach to support the co-evolution of
  object-oriented design and implementation}. PhD thesis. PhD thesis, Vrije
  Universiteit Brussel, ;  2001.

\bibitem{DEMIMA}
Guéhéneuc YG, Antoniol G. DeMIMA: A Multilayered Approach for Design Pattern
  Identification. {\it IEEE Transactions on Software Engineering} 2008\string;
  34(5)\string: 667-684.
\newblock \href {\doibase 10.1109/TSE.2008.48} {doi: 10.1109/TSE.2008.48}

\end{thebibliography}

\end{document}